%% file: main.tex
\documentclass{IEEE-Con-Sys-mag}

\jvol{XX}
\jnum{XX}
\paper{8}
\jmonth{June}
\jname{IEEE CONTROL SYSTEMS}
\pubyear{2020}
\usepackage[compress]{cite}
\usepackage{amssymb,subfig}
\usepackage[font=footnotesize]{caption}
\usepackage{tikz,pgfplots}
\usepackage{graphicx}
\usepgfplotslibrary{fillbetween}
\newcommand{\vect}[1]{\boldsymbol{#1}}

\pgfplotsset{ 
  compat=newest, 
   legend style =
  {font=\footnotesize },
  label style = {font=\footnotesize},
every tick label/.append style={font=\scriptsize}
  }

\begin{document}

\title{Awareness in collective decision-making\stitle{Modeling and control in a game-theoretic framework}}

\author{{M}ENGBIN YE, LORENZO ZINO, and MING CAO}
\affil{}

\maketitle

\dois{}{}

\begin{summary}
\summaryinitial{F}or a society to remain healthy and prosperous, people must collectively behave and act to contribute to the common good, even if there is often a tradeoff against their individual benefit. Paradigmatic examples include the adoption of sustainable behaviors and technologies to combat the climate crisis, and the mobilization for collective action to promote the rights and freedoms of repressed minorities. In this tutorial, we illustrate how game theory and network systems theory can be powerful tools to model and study this collective decision-making problem. We provide examples of how awareness of this tradeoff can impact collective change toward the societal good, exploring   different problem contexts such as sustainable behavior and collective action. Finally, we review recent developments using systems and control-theoretic approaches to generate awareness and guide the emergent population dynamics towards a desired outcome, and conclude by highlighting new research and application frontiers.
\end{summary}

\vspace{11pt}

\chapterinitial{T}hroughout history, a cornerstone of a healthy and prosperous society is the willingness of its population to adopt actions and behaviors that contribute to the collective good, rather than those that maximize individual benefits. This is because in many situations there is an inherent and fundamental tradeoff between actions or behaviors that provide greater individual benefit, and those that are more costly to the individual but can generate a significant reward for the entire population (that is, a collective good) if enough people do so collectively. When examining the issue from an individual's perspective, intuitive assumptions about the selfish nature of people (which extend to animals and other species) will naturally lead to the conclusion that a society is unlikely to collectively adopt the actions or behaviors that maximizes the collective good. Indeed, key examples throughout history have highlighted the negative impacts that can ripple across society when this occurs. For instance, the excessive use of antibiotics, which has immediate beneficial effects for the individual, has resulted in the emergence of antibiotic-resistant bacterial pathogens~\cite{Davies2010}. Overfishing in the Atlantic northwest from the 1950s onward, which provides immediate benefits for individual fishermen, has led to the depletion of fish stocks and, ultimately, the collapse of entire fishing industries~\cite{Myers1997}. However, there is perhaps even more evidence that a society can act towards the collective good, leading to sustained prosperity and growth. An example is the volunteer blood donation system, whereby in several countries the blood used for transfusions to treat sick or injured patients comes from volunteers who contribute to the well-functioning of the healthcare system, despite receiving no immediate benefits for their selfless action~\cite{Culyer1973}.

Our world today is faced with an ever increasing number of societal threats and challenges, including the climate crisis, pandemics, democratic backsliding, and the transition to clean and renewable energy. As we discovered during COVID-19, a successful pandemic response requires not only medicinal advances but society's collective adoption of self-testing, self-protective behaviors, and vaccines, which are costly and inconvenient for each individual~\cite{Bavel2020}. The transition to clean and renewable energy and other pro-environmental behaviors will bring about obvious benefits for society at large, but will require concerted effort by each person to forgo and change from existing, individually beneficial, behaviors and technologies~\cite{Welch2018,Bowen2017}. More examples are discussed in details in ``\nameref{side:dilemma}."

\begin{pullquote}
There is a need, now more than ever, to understand this fundamental tension and tradeoff between the collective good and individual benefit\end{pullquote}

Considering these challenges, there is a need, now more than ever, to understand this fundamental tension and tradeoff between the collective good and individual benefit. This will allow policymakers to develop effective and efficient strategies to overcome the adaptiveness and selfishness of people favoring individual benefits, so as to influence and steer a population towards a collective decision-making outcome that favors the collective good. Unlike engineered systems, the policies and strategies that are considered applies to people, and there are important ethical and practical considerations that restrict the possible levers of influence, necessitating more sophisticated modeling and analysis. And yet, in agreement with the Road Map 2030 of the IEEE Control Systems Society~\cite{2030}, we believe that the systems and control community can offer important contributions to address such challenges by combining classical approaches with emerging methods, including techniques from other disciplines.

\begin{sidebar}{Tradeoff between collective good and individual benefit}
\section[Tradeoff between collective good and individual benefit]{}\label{side:dilemma}

\sdbarinitial{T}he tension and tradeoff between collective good and individual benefit can manifest itself in different ways depending on the scenario. Here, we briefly discuss two examples.

\subsection{Protest movements and collective action}
    In the context of political and social science, collective action is when a group of people coordinate together and carry out activities or actions to achieve a common objective, often concerning political or societal change~\cite{thomas_2022_SA}. This can range from lobbying politicians and voting in democratic processes, to participating in protest movements~\cite{moghaddam_2016_SA}. Intuitively, participation in collective action comes with an individual cost, which can be the energy and effort expended to lobby or vote, or on a more extreme end of the spectrum, imprisonment or physical harm when protesting against authoritative regimes. In other words, an individual benefits from not participating in collective action, if they think that the objective will still be achieved because enough others participate in the collective action. This is an example of the classical free-rider problem that is studied in both biology and sociology~\cite{olson_1965_SA,nowak_2006_SA}. If enough people participate in collective action, then social or political change may occur, which in many instances brings about a collective good for either the population, or a repressed minority. Examples include the women's suffrage movement in the 20th Century, the American Civil Rights Movement of the 1960s, marriage equality, and the Black Lives Matter movement in the United States.

\subsection{Social norms and conventions}
Social norms and conventions (norms for short) are an integral part of society and culture, providing guidelines on how people should interact and behave~\cite{Young2015_SA}. Norms can vary from harmless and inconsequential (for instance, the use of American or British spelling in academic writing) to having substantial impact on the individual and the society (for example, norms on footbinding for women in rural China~\cite{Young2015_SA}, or norms on cycling and using public transportation for commuting~\cite{sparkmann_2020_SA}). While exact definitions vary, most norms share the same set of characteristics. First, there are at least two options or choices which are alternatives of one another, for instance, using public transport versus driving to work. Second, a benefit or improvement is realized precisely when people coordinate to adopt the same option, for instance, everyone using public transport reduces overall traffic congestion. Third, it is possible that across a large population, there is strong local conformity and global diversity in the behavior that is normative~\cite{Young2015_SA}. 
For social norms, a tradeoff between individual benefit and collective good can arise if a population is currently broadly adopting an option that is inferior to an alternative. In this situation, there would be an overall increase in the collective good if everyone switched to the superior alternative, but each individual would (initially) incur a cost to switch away from the status quo, since benefit is obtained from coordinating on the same option.

\end{sidebar}

\section{Article content and organization}

This article aims to provide a tutorial on how to model human behavior in collective decision-making scenarios that involve a tradeoff between the collective good and individual benefit. Across the next two sections, we first give some motivating examples of collective decision-making scenarios in which there is a tradeoff between the collective good and individual benefit, before introducing evolutionary game theory as a flexible and powerful modeling framework for these scenarios. 
The main application is the paradigm of collective change, whereby we consider the shifting of a population from widespread adoption of one behavior or action to another that provides a greater collective good, and examine strategies to promote collective change that center around increasing awareness of desired behaviors within the population. 
Next, we build on systems-theoretic models of opinion dynamics to introduce a game-theoretic model that allows us to study the coevolution of actions and opinions. Finally, we illustrate how raising awareness through opinion dynamics, and the subsequent coevolution of actions and opinions, is another pathway to collective change. We conclude by discussing perspectives and key future directions.


\section{What's in a game?}


Consider a population of $n\geq 2$ players, indexed by the set $\mathcal{V} = \{1,2,\hdots, n\}$. Player~$i$ can select a strategy from the strategy set $\mathcal A_i$, the elements of which can be discrete or continuous depending on the particular game, and they receive a payoff determined by the payoff function $f_i : \mathcal A \to \mathbb R$. Here, $\mathcal A = \mathcal A_1 \times \hdots \times \mathcal A_n$ is the set of strategies for all players, and we can similarly define $\vect f = [f_1, \hdots, f_n]^\top$ as the payoff vector function. If one defines $z_i \in \mathcal A_i$ as the strategy selected by player~$i$, then the strategy profile of the population can be written as $\vect z = [z_1, \hdots, z_n]^\top \in \mathcal A$. A key concept in game theory is that the payoff player~$i$ receives depends on the strategy of other players, and thus we write $f_i(z_i, \vect z_{-i})$ to denote the payoff for selecting strategy $z_i \in \mathcal A_i$ when the strategy profile for other players is captured by the vector $\vect z_{-i} = [z_1, \hdots, z_{i-1}, z_{i+1}, \hdots, z_n]^\top$. With these variables, we can then define a game as $\Gamma = (\mathcal{V}, \mathcal{A}, \vect f)$. 
Two central concepts in game theory are that of Nash equilibrium and best response.

\begin{definition}[Best response]\label{def:BR}
    Given game $\Gamma = (\mathcal{V}, \mathcal A, \vect f)$ and a strategy profile $\vect{z}\in\mathcal A$, the set of best response strategies for player $i$ is  
    \begin{equation}\label{eq:BR_def}
    \mathfrak{B}_i (f_i(\cdot, \vect z_{-i})) :=\text{argmax}_{s\in\mathcal A_i} f_i(s, \vect z_{-i}).
\end{equation}
\end{definition}

\begin{definition}[Nash equilibrium]\label{def:NE}
    A strategy profile $\vect z^*$ is a Nash equilibrium for the game $\Gamma = (\mathcal{V}, \mathcal A, \vect f)$ if there holds 
       $z_i^*\in  \mathfrak{B}_i (f_i(\cdot, \vect z^*_{-i}))$, for all $i\in\mathcal V$.
\end{definition}

In other words, a Nash equilibrium of a game is a strategy profile from which no player would benefit (increase their payoff) by unilaterally deviating from said profile, since there holds
        $f_i(z_i^*, \vect z_{-i}^*) \geq f_i(s, \vect z_{-i}^*)$ for all $s \in \mathcal{A}_i$.

Classic game theory provides a static description of a ``game" in terms of the characterization of its Nash equilibria, under the assumption players do not revise their strategy after selection. In the context of modeling decision-making, we are interested in individuals who dynamically revise their strategy. To describe such scenarios, evolutionary game theory was developed to extend classical game theory~\cite{Sandholm2010}, whose basic formalism and concepts are described in ``\nameref{side:egt}."

In the following, we present two classes of games that are used to capture key aspects of human behavior: Network coordination games and public goods games.

\begin{sidebar}{Evolutionary game theory}
\section[Evolutionary game theory]{}\label{side:egt}
    \setcounter{sequation}{0}
    \renewcommand{\thesequation}{S\arabic{sequation}}
    \setcounter{stable}{0}
    \renewcommand{\thestable}{S\arabic{stable}}
    \setcounter{sfigure}{0}
    \renewcommand{\thesfigure}{S\arabic{sfigure}}

\sdbarinitial{E}volutionary game theory is an extension of classical game theory, with a focus on how strategies among multiple players evolve through repeated interactions~\cite{hofbauer1998_SA,riehl_2018_SA}. Here, we present the general mathematical formulation, which we then adapt to a range of games throughout this article for capturing specific interaction scenarios. Broadly speaking, evolutionary games involve a set of players who are making decisions on which strategies to select. This decision centers on a payoff function, that defines the reward the game. The game is played repeatedly as players revise their strategy following a revision protocol. Interactions often constrained over a network, and not every player may be active at each point in time. 


\subsection{Strategy revision}
Consider a population of $n\geq 2$ players, indexed by the set $\mathcal{V} = \{1,2,\hdots, n\}$, who play a game  $\Gamma = (\mathcal{V}, \mathcal{A}, \vect f)$, with $\mathcal A=\mathcal A_1\times\dots\times \mathcal A_n$ and $\vect{f}=[f_1,\dots,f_n]^\top$, where $\mathcal A_i$ and $f_i$ are the set of strategies and the payoff function for player $i$, respectively. Players interact repeatedly over time and are given the chance to revise their strategy, leading to evolution of the strategy profile $\vect z(t)$. To ensure consistency, all games in this article are played over discrete time steps $t = 0, 1, 2, \hdots$, but note that continuous-time formulations exist for many of the games discussed. 

For a given time step $t\in\mathbb N$, we suppose that a subset $\mathcal{R}(t) \subseteq \mathcal V$ of players becomes active and makes a decision on which strategy to select for time $t+1$. Namely, we write
\begin{sequation}
    z_i(t+1) = \left\{\begin{array}{ll}F(\vect z(t))\,,\quad &\text{if } i \in \mathcal R(t),\\z_j(t+1) = z_j(t)&\text{otherwise.}\end{array}\right.
\end{sequation}
Here, $F(\cdot): \mathcal A \to \mathcal A_i$ is termed the revision protocol, which we detail shortly. The assumption $\mathcal R(t) = \mathcal V$ and $\vert \mathcal R(t) \vert = 1$ for all $t$ leads to the so-called synchronous and asynchronous updating dynamics, respectively. For non-synchronous updating, different assumptions can be imposed on how the sequence $\mathcal R(0), \mathcal R(1),\hdots$ is generated, including both deterministic and stochastic mechanisms.

The revision protocol $F$ defines the decision-making process that player~$i$ employs to select their strategy. As  observed in experimental social psychology~\cite{Mas2016_SA}, people typically tend to maximize their payoff in a myopic fashion. This assumption lead to the definition of the  myopic best-response revision protocol, which is defined as $F(\vect z(t)) \in \mathfrak{B}_i (f_i(\cdot,\vect z_{-i}(t)))$, or 
\begin{sequation}\label{eq:BR_generic}
    z_i(t+1) \in  \mathfrak{B}_i (f_i(\cdot, \vect z_{-i}(t))),
\end{sequation}
that is, assuming that when they revise, individuals always select the strategy that maximizes their current payoff. Note that $\mathfrak{B}_i (f_i(\cdot,\vect z_{-i}(t)))$ may contain more than a single element (that is, there are multiple strategies which yield the highest payoff). In this case, simple tie-break rules can be defined a priori, such as having an ordered preference of the elements in $\mathcal A_i$, selecting uniformly at random from the elements in $ \mathfrak{B}_i (f_i(\cdot, \vect z_{-i}(t)))$, or $z_i(t+1) = z_i(t)$ if $z_i(t) \in  \mathfrak{B}_i (f_i(\cdot, \vect z_{-i}(t)))$.

Other revision protocols include imitation, relative best-response, and noisy best-response. The latter is motivated by the observation that players (especially people) do not always select a strategy that maximizes payoff, whether it be due to making a mistake or personal variation. One of the most popular implementations is known as loglinear learning or logit updating~\cite{blume1995_SA}. Namely, the probability of selecting strategy $s\in \mathcal A_i$ is
\begin{sequation}\label{eq:loglinear}
    \mathbb P [z_i(t+1) = s | \vect z(t)] = \frac{e^{-\sigma_i f_i(s,\vect z_{-i}(t))}}{\sum_{r\in \mathcal A_i}e^{-\sigma_i f_i(r,\vect z_{-i}(t))}},
\end{sequation}
where $\sigma_i \in [0,\infty)$ is the bounded rationality of player~$i$. Namely, $\sigma_i = 0$ means player~$i$ selects strategies uniformly at random, while $\sigma_i \to \infty$ yields \eqref{eq:BR_generic}. 

Finally, the myopic best-response protocol establishes a connection between the Nash equilibrium of the game $\Gamma$ and the dynamics induced by \eqref{eq:BR_generic} as individuals repeatedly interact on $\mathcal G$. 
In fact, considering the definition in Definition~\ref{def:NE} and \eqref{eq:BR_generic}, we observe that 
    a state $\vect z^*$ is an equilibrium of the best-response dynamics, that is, $z_i^*(t+1) = z_i^*(t)$ for all $i\in\mathcal V$ and $t$, if and only if $\vect z^*$ is a Nash equilibrium of the game $\Gamma$.

Evolutionary game theory is a general modeling framework that is relevant in a wide range of application domains~\cite{quijano2017_SA,Traulsen2023_SA}. In this article, we focus on games and models which involve humans (people) as players, interacting over social networks, with games that center around collective decision-making and tradeoffs between the collective good and individual benefit.

\end{sidebar}

\subsubsection{Network coordination games}

\begin{figure*}
    \centering
\subfloat[Payoff mechanism]{\input{fig/coord} \label{fig:coordination}}
\subfloat[Nash equilibria]{\input{fig/coord_ne} \label{fig:nash}}
\caption{Binary network coordination game. In (a), we illustrate the mechanisms of the payoff function. In (b), we consider a network with $6$ players that interact on $2$ triangular cliques with a link between the two cliques. There are up to four distinct Nash equilibria of the game, which include not only the two consensus strategy profiles (left column), but also two nontrivial configurations with local coordination but global diversity (right column). These two nontrivial configurations are Nash equilibria if and only if $\alpha\in[-\frac12,1]$, verified by evaluating the payoff function of the two individuals that bridge the two cliques. Otherwise, only the two consensus strategy profiles are Nash equilibria. }
\end{figure*}
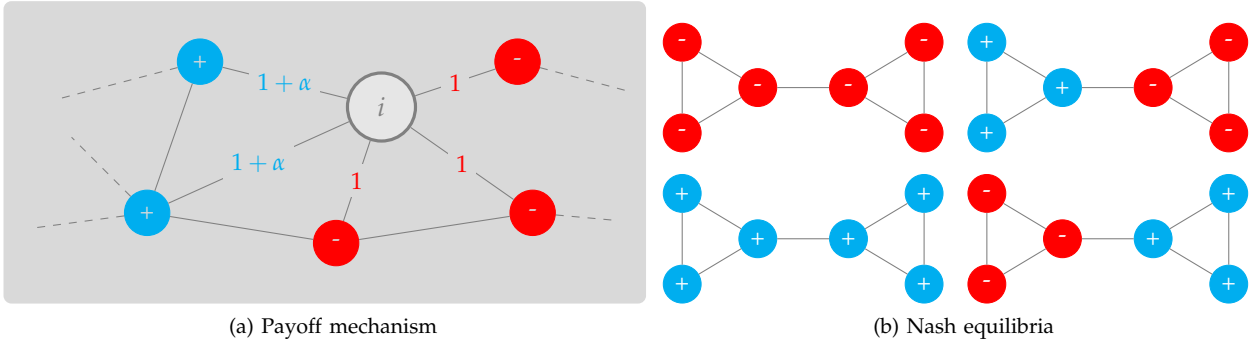

We start by describing the simplest scenario, involving two players, and assume they share the same strategy set $\mathcal A$. Coordination games are a class of matrix games~\cite{riehl2018survey}, for which the payoff  that player $i$ would receive is captured by a $|\mathcal A|\times |\mathcal A|$ matrix $M^{(i)}$. Specifically, the generic entry $m^{(i)}_{ab}$ represents the payoff that player $i$ receives for selecting strategy $a$ against a player $j$ selecting $b$, that is, $f_i(a,b)=m^{(i)}_{ab}$. 
We assume that all players share the same payoff matrix, denoted as $M$.

This 2-player game structure is straightforwardly extended to a network scenario, whereby players interact on a (weighted) network $\mathcal G=(\mathcal V,\mathcal E, A)$, with $A$ the (weighted) adjacency matrix such that $a_{ij}>0$ if and only if $i$ interacts with $j$, and $a_{ij}$ captures the strength of the interaction.  Without any loss in generality, we can assume that $A$ is stochastic, that is, $A\vect{1}=\vect{1}$, where $\vect1$ is the $n$-dimensional all-$1$ vector. 
In a network game, each player plays one instance of the $2$-player game against each of their neighbors, with the overall payoff obtained by summing over all instances of the game, weighted by the strength of the corresponding interaction~\cite{Jackson2015}. Thus, the reward that player $i$ receives from selecting action $a\in\mathcal A$ can be written as
\begin{equation}\label{eq:payoff_network}
     f_i(a,\vect z_{-i})=\sum\nolimits_{j\in\mathcal V}a_{ij}m_{az_j}.
\end{equation}

In a (pure) network coordination game, a player receives a nonzero, strictly positive payoff if and only if they select the same strategy as their neighbor. In other words, matrix $M$ is a positive definite diagonal matrix. Coordination games on networks is a classical paradigm for collective decision-making mechanisms, including for modeling social norms and conventions~\cite{peytonyoung2015social_norms,Ye_2025_chapter_norms}. This is because this paradigm centers on scenarios in which there is advantage for people to adopt the same strategy, or to capture social pressure and the tendency to conform~\cite{young2011dynamics,ye2021nat}. Below, we use a binary strategy realization to illustrate how we can model the diffusion of innovation, or the evolution of social norms, since adopting the same strategy as your neighbors yields several direct and indirect advantages.

\begin{example}[Binary network coordination game]\label{ex:coordination}
For a network coordination game with $\mathcal A=\{-1,+1\}$, the payoff matrix can always be re-scaled to read as
\begin{equation}\label{eq:coordination}
    M=\begin{bmatrix}
        1&0\\0&1+\alpha
    \end{bmatrix},
\end{equation}
where the parameter $\alpha>-1$ captures a relative advantage (if positive) or disadvantage (if negative) of strategy ``+1" with respect to strategy ``-1." In other words, individual $i$ receives a unitary payoff from coordinating with a neighbor on the ``-1" strategy, while they receive payoff $1+\alpha$ if they coordinate on the ``+1" strategy.

Embedding the game on a network, and inserting \eqref{eq:coordination} into \eqref{eq:payoff_network}, we obtain 
\begin{subequations}\label{eq:payoff_coordination}
\begin{align}
   &f_i(+1,\vect{z}_{-i})=\frac12(1+\alpha)\sum\nolimits_{j\in\mathcal V} a_{ij}(1+z_i),\\
   &f_i(-1,\vect{z}_{-i})=\frac12\sum\nolimits_{j\in\mathcal V} a_{ij}(1-z_j),
\end{align}
which can also be compactly written, for $s\in\{-1,+1\}$, as 
\begin{equation}
   f_i(s,\vect{z}_{-i})=\frac14\sum_{j\in\mathcal V} a_{ij}\Big( (1+\alpha)(1+s)(1+z_i)+(1-s)(1-z_j)\Big).
\end{equation}
\end{subequations}

In the context of a 2-player game, a coordination game has just two Nash equilibria, coinciding with both players selecting the same strategy (as can be checked by using Definition~\ref{def:NE}). On a network however, a coordination game can have multiple equilibria, shaped by the structure of the network and by the relative advantage  $\alpha$. Consider the example in Fig.~\ref{fig:coordination}, under the assumption that all edge weights are equal. Notice that node $i$ is interacts with two ``+1" and three ``-1" neighbors, which means that $+1\in\mathfrak{B}_i (f_i(\cdot, \vect z_{-i}))$ if and only if $2+2\alpha\geq 3$, that is, if $\alpha\geq 1/2$. 
\end{example}

From these local interactions, nontrivial collective behavior can emerge. In fact, while $\vect z = -\vect 1$ and $\vect z = +\vect 1$ consensus strategy profiles are always Nash equilibria, a network coordination game may possess other nontrivial Nash equilibria, in which the population is partitioned into clusters. The existence of these equilibria depends on the network. In particular, assume that there is a set of nodes  $\mathcal S\subset \mathcal V$  such that
$\sum_{j\in\mathcal S}a_{ij}\geq \frac{1}{2+\alpha}$ for all $i\in\mathcal S$, which is said to be cohesive~\cite{morris2000contagion,Arditti2021}. Then, a best-response strategy for all nodes within the set $\mathcal S$ is to collectively coordinate on strategy $1$, since $f_i(+1,\vect{z}_{-i})>f_i(-1,\vect{z}_{-i})$, for all $i\in\mathcal S$. This implies that whether there exists a Nash equilibrium in which the nodes in $\mathcal S$ all coordinate on strategy $+1$ is independent on the strategy of the nodes not in $\mathcal S$. A similar property would hold if the nodes coordinate on strategy $-1$, with the requirement that $\sum_{j\in\mathcal S}a_{ij}\geq \frac{1+\alpha}{2+\alpha}$, due to the bias in the relative advantage. If the network is such that multiple sets hold this property, there are  Nash equilibria with local coordination (within each set) and global diversity, as illustrated in Fig.~\ref{fig:nash}. 


\subsubsection{Public goods games}

Public goods games are a paradigm used to study social dilemmas, that is, contexts in which individuals have to make decisions between selfish choices which provide higher personal payoffs and altruistic choices, which yield superior collective good~\cite{Bramoull2007}. This is the case of many real-world scenarios such as joining protest movements and collective action~\cite{olson_1965_collective_action_public_goods}, discussed in ``\nameref{side:dilemma}."

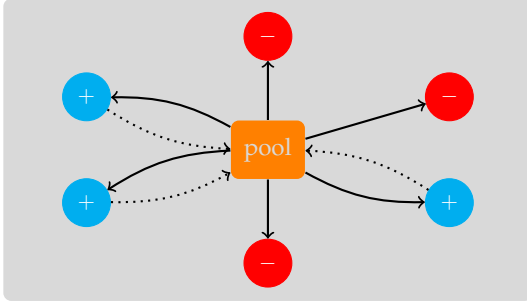
\begin{figure}
    \centering
      \input{fig/pgg2}
    \caption{Payoff mechanism of a public goods game. Three of the six players give a unit contribution to the public pool (dotted arrows), and all players receive an equal split of the reward (solid arrows), that is,  $\rho(\frac12)$, yielding the payoff function in \eqref{eq:payoff_pgg_general}. }
    \label{fig:pgg}
\end{figure}

There are many different formalizations, but in the simplest scenario, each individual in the population $\mathcal V=\{1,\dots,n\}$  selects from a binary strategy set, deciding whether they ``defect" (denoted by $-1$) or ``cooperate" (denoted by $+1$), which represent the selfish and altruistic choice, respectively. 
If player~$i$ decides to cooperate, they contribute a unit amount to the public goods pool ---this assumption can be relaxed, obtaining more complex models where the player can decide how much to contribute (if at all)~\cite{Bramoull2007}. Regardless of their individual decision, each player receives a reward that depends on the size of the public goods pool, that is, on the strategy of the others and in particular on how many players have cooperated. By denoting 
\begin{equation}\label{eq:zeta}\zeta=\frac{|\{i\in\mathcal V:z_i=+1\}|}{n}=\frac{n+\sum_{i\in\mathcal V} z_i}{2n}\end{equation}
as the fraction of cooperators, and assuming the pool is divided evenly across all players, the payoff functions for cooperation and defection are
\begin{subequations}\label{eq:payoff_pgg_general}
\begin{align}
   &f_i(+1,\vect{z}_{-i})=\rho\Big(\frac{n+1+\sum_{i\in\mathcal V\setminus\{i\}} z_i}{2n}\Big)-1,\\
   &f_i(-1,\vect{z}_{-i})=\rho\Big(\frac{n-1+\sum_{i\in\mathcal V\setminus\{i\}} z_i}{2n}\Big).\end{align}
\end{subequations}
The function $\rho(\zeta):[0,1]\to\mathbb R_{\geq 0}$ describes how the public goods pool is impacted by contributions from cooperators. It is typically monotonically increasing, and has certain consistent characteristics that capture the social dilemma. For instance, the pool can be proportional to the amount of people who cooperate, or there could be a threshold, whereby the pool is available for sharing only if a critical mass of cooperators has been reached. The general mechanism is illustrated in Fig.~\ref{fig:pgg}. Of particular interest due to its immediate interpretation and analytical tractability is when $\rho$ is linear, as discussed below.

\begin{example}[Linear-Reward Public Goods Game]\label{ex:pgg}
When the pool grows linearly as cooperation increases and starts from zero, that is, $\rho(\zeta)=r\zeta$, 
the payoff functions in \eqref{eq:payoff_pgg_general} 
can be compactly written, for $s\in\{-1,+1\}$, as 
\begin{equation}\label{eq:payoff_pgg}
   f_i(s,\vect{z}_{-i})=r\Big(\frac{n-1+\sum_{j\in\mathcal V\setminus\{i\}} z_j}{2n}\Big)+\Big(\frac{r}{n}-1\Big)\frac{s+1}{2}.
\end{equation}
Here, the constant $r$ represents the public goods multiplier~\cite{Bramoull2007,govaert_rationality_2021}, that is, the increased societal value obtained thanks to cooperation. It is reasonable to assume $1<r<n$. On the one hand, $r<1$ would imply that cooperation does not provide any societal benefit. On the other hand, $r>n$ would imply that it is always beneficial to cooperate. In both cases, there is no social dilemma. For $1<r<n$, instead, an individual's best response is always to defect, since $f_i(-1,\vect z_i)>f_i(+1,\vect z_i)$ for all $i\in\mathcal V$ and $\vect z_i\in\{-1,+1\}^{n-1}$. This in turn implies that the only Nash equilibrium of the game is the all-defecting Nash equilibrium. However, this equilibrium is suboptimal not only from the societal level, but also from the perspective of any single player. In fact, in the all-defecting Nash equilibrium, all individuals receives payoff equal to $f_i(-1,-\vect1)=0$, while in the all-cooperation state, each individual would receive $f_i(+1,+\vect1)=r-1>0$. 
\end{example}

\section{Collective change}

Having formally introduced evolutionary game theory as a modeling framework (see ``\nameref{side:egt}"), and coordination games and public goods games as specific games of interest (see Examples~\ref{ex:coordination}--\ref{ex:pgg}), we now introduce the paradigm of collective change, formalizing the tradeoff between collective good and individual benefit within a unified problem setting. To simplify our exposition, and consistent with the examples introduced above, we consider games in which there are precisely two strategies, $\mathcal A_i = \{-1,+1\}$. In this scenario, we consider a population of individuals, where a generic individual $i$ can adopt at a generic time $t$ one of the two strategies, denoted as $x_i(t)=-1$ or $x_i(t)=+1$, respectively, with $t\in\mathbb N$. We shall refer to strategy $-1$ as the status quo in coordination games or defection in public goods games, and to $+1$ as innovation or cooperation, respectively. Extending \eqref{eq:zeta}, the fraction of adopters of action $+1$ at time $t$ can be conveniently denoted as
\begin{equation}
    \label{eq:zeta_t}
    \zeta(t)=\frac{|\{i\in\mathcal V:x_i(t)=+1\}|}{n}=\frac12+\frac{1}{2n}\sum\nolimits_{i\in\mathcal V}x_i(t).
\end{equation}
Using the formalism of evolutionary game theory, the individuals play the game and revise their strategy according to some strategy revisions such as the myopic best-response revision protocol in \eqref{eq:BR_generic} or a logit in \eqref{eq:loglinear}, previously described. For the sake of readability, we report all the variables and parameters of the two classes of games considered in this article in Table~\eqref{tab:game}.  In this scenario, we can define the following paradigm.


\begin{definition}[Collective Change]\label{def:collective_change}
Consider a population $\mathcal V=\{1,\dots,n\}$ of individuals repeatedly revising their strategies for a given evolutionary game $\Gamma = (\mathcal V, \mathcal A, \vect f)$, with the population strategy profile $\vect z(t)$. Suppose that the population is initially at a status quo of consensus on the $-1$ strategy, $\vect{z}(0)=-\vect 1$ (or, equivalently, $\zeta(0)= 0$).
The collective change problem consists of identifying the conditions (for instance, on the game, the network, the potential inclusion of controlled players) such that the population asymptotically converges to a new consensus state on the $+1$ strategy. We say collective change is achieved if $\lim_{t\to\infty}\vect z(t) = +\vect 1$ (or, equivalently,  $\lim_{t\to\infty}\zeta(t)= 1$).
\end{definition}

In this article, we assume that collective change is desirable, and if a policymaker is considering implementing interventions or controls on the population, then it is to facilitate and promote collective change. This perspective covers a great number of scenarios, such as promoting the adoption of pro-environmental behavior or technologies~\cite{Hoffmann2024}, collective action on progressive issues~\cite{thomas_2022_mobilise}, adoption of vaccines or self-protective behaviors during epidemic outbreaks~\cite{Bavel2020}. Naturally, limiting or restricting collective change can also be considered, such as ensuring the broader population continues to participate in democratic processes and activities~\cite{waldner_2018_democratic_backslide}, or limiting the adoption of violent tactics in protest movements~\cite{thomas_2022_mobilise}. To maintain a clear exposition, we focus on collective change that is desirable.

\begin{table}
\center
\caption{Variables and parameters of the games.\label{tab:game}}
\begin{tabular}{|c|c|}
\hline
Symbol& Meaning \\
\hline
$x_i(t)$& Action of player $i\in\mathcal V$ at time $t\in\mathbb N$\\
$\zeta(t)$& Fraction of adopters of action $+1$ at time $t\in\mathbb N$\\
$\mathcal R(t)$& Players who revise action at time $t\in\mathbb N$\\
$a_{ij}$& Strength of interaction between player $i$ and $j$\\
$n$& Number of players in the population\\
$\alpha$& Relative advantage of $+1$ in coordination games\\
$r$& Public good multiplier in public goods games\\
\hline
\end{tabular}
\end{table}

\subsection{Collective change for coordination games}

\begin{figure*}
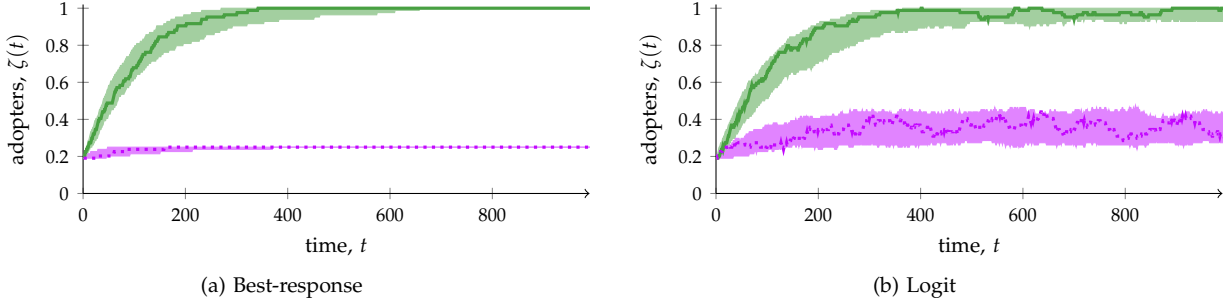

    \centering
    \subfloat[Best-response\label{fig:diffusion1}]{\input{fig/coord1}}\,\,
    \subfloat[Logit\label{fig:diffusion2}]{\input{fig/coord2}}
    \caption{Collective change for coordination games. Simulations performed on a network with $n=84$ nodes, reconstructed from real-world interactions~\cite{malawi_net}. We report one illustrative simulation for each scenario, while the colored area represents the 95\% simulations envelope (over 100 independent realizations of the processes). Panel (a) shows two trajectories for the best-response dynamics with 16 players acting as committed minority, but with different placement in the network. The dotted violet curve and solid green curve show outcomes when committed minority are placed at nodes with lowest eigenvalue centrality and highest eigenvector centrality, respectively. For the former, only a few players join the committed minority, while collective change is achieved for the latter. In (b), we compare two trajectories for the logit dynamics in \eqref{eq:loglinear} with $\sigma_i=4$ for all $i\in\mathcal V$, placing $16$ committed minority players at nodes with largest eigenvector centrality. We vary the relative advantage, with $\alpha=0$ for the dotted violet curve and $\alpha=1$ for the green solid cure. The latter displays `fast' collective change, which is not observed in the former within the simulation time horizon. }
    \label{fig:diffusion}
\end{figure*}

Assuming that the population follows the myopic best-response revision protocol in \eqref{eq:BR_generic} for a network coordination game, from the payoff functions in \eqref{eq:payoff_coordination} we can explicitly write the dynamics for active player $i\in \mathcal R(t)$ as
\begin{equation}\label{eq:coord_BR}
    x_i(t+1)=\left\{\begin{array}{ll}+1&\text{if }\sum_{j\in\mathcal V}a_{ij}x_j(t)>-\frac{\alpha}{2+\alpha},\\
    -1&\text{if }\sum_{j\in\mathcal V}a_{ij}x_j(t)<-\frac{\alpha}{2+\alpha},
    \end{array}\right.
\end{equation}
with $x_i(t+1)=x_i(t)$ if $\sum_{j\in\mathcal V}a_{ij}x_j(t)=-\frac{\alpha}{2+\alpha}$.

Here, Definition~\ref{def:collective_change} is analogous to the widely studied innovation diffusion, social diffusion, and social learning problems~\cite{montanari2010spread_innovation,young2011dynamics,kreindler2014rapid_diffusion}. When $\alpha > 0$, the players receive a greater collective good if they are at the state $\vect z = +\vect 1$ (consensus on the innovation), compared to the state $\vect z = -\vect 1$ (consensus on the status quo). 
However, when starting from $\vect z=-1$, no individual is willing to unilaterally change their strategy as they would receive a lower payoff by adopting $+1$ when all their neighbors are still adopting $-1$, as per \eqref{eq:coord_BR}, making it impossible to achieve collective change in the absence of an external intervention.

There is an established literature on how to achieve collective change for coordination games on networks. 
The typical approach used involves the introduction of a committed minority (zealots, innovators, opinion leaders) who select the strategy $+1$; such an introduction can reflect the deliberate attempt by a policymaker to promote collective change, or the natural emergence of a minority within the population who seek change. Studies have investigated the size of committed minority required, identifying that there is often a threshold value above which fast collective change is achieved~\cite{centola2018experimental_tipping}. This has been further extended to consider optimal placement of committed minority over the network~\cite{morris2000contagion,kempe2003_maxspread,Acemoglu2011}. This problem is, in general, NP-hard ---but efficient heuristics have been proposed~\cite{Como2022supermodular}. Figure~\ref{fig:diffusion1} illustrates how collective change can be achieved in a network coordination game by a thoughtful placement of the committed minority. Moreover, from a policymaker's perspective, we can ask whether increasing the incentives of the innovation  facilitates collective change~\cite{Riehl2016,riehl2018incentive,Liu2026}. Most often, this is equivalent to changing the payoff structure of the game, which in \eqref{eq:coordination} is equivalent to adjusting $\alpha$. This can either be a one-shot intervention at $t = 0$, or implemented dynamically over time, and either target all players or a carefully selected subset~\cite{MartinezPiazuelo2022b,Zino2025tac}.

Following a different approach, if we assume that players use the logit updating mechanism described in \eqref{eq:loglinear}, then it is necessary to consider a slightly different definition of collective change, since the stochasticity and ergodicity of the dynamics implies that state $\vect z=+1$ is always reached with probability 1, but the some small oscillations are possible (see the trajectories in Fig.~\ref{fig:diffusion2}). In this case, it is possible to identify conditions on $\sigma$, $\alpha$, and the network structure such that $\vect z(t) \to +\vect 1$ occurs rapidly independent of the number of players. Hence,  the question is slightly re-framed to determine whether $\bar t$ scales greater than linearly with $n$. Roughly speaking, `fast' collective change occurs if $\sigma$ is sufficiently small, or $\alpha$ is sufficiently large. Intuitively, favorable conditions for collective change include when individuals are willing to adopt choices that do not maximize their individual benefit (low $\sigma$) or when the innovation offers a substantial advantage over the status quo. This can be observed by comparing the two trajectories in Fig.~\ref{fig:diffusion2}. More details can be found in~\cite{montanari2010spread_innovation,young2011dynamics,kreindler2014rapid_diffusion}. We can of course consider the role of a committed minority in the presence of logit update~\cite{ye2021nat,gao_2024_diffusion_networks}.

\subsection{Collective change for public goods games}

As discussed above, for the public goods game with $1<r<n$, the only Nash equilibrium is for everyone to defect, that is, $\vect z = -\vect 1$. It is not difficult to show that, assuming individuals adopt a best-response updating mechanism, then for all $\vect z(0)$, there holds $\lim_{t\to\infty} z(t) = -\vect 1$ irrespective of $r$ or $n$. The tradeoff between collective good and individual benefit is understood as follows. If $\vect z(t) = +\vect 1$, then the entire population receives the full public good with multiplier $r$, but any player who defects would receive a greater benefit because they would receive their shares of the public good (due to others cooperating) without having to contribute. As a result, collective change cannot be achieved by allowing players to repeatedly revise their strategies. Existing literature has examined the use of different approaches to promote cooperation, such as peer punishment or incentives~\cite{Fehr2002,Gachter2008,Rand2009,Zhu2020,Sun2023}, or relying on the incorporation of reputation-based mechanisms~\cite{Milinski2002,Elokda2023}. Interestingly, similar to the committed minority driven diffusion in coordination games, some of these methodologies have been proposed in combination with the introduction of a committed minority of cooperators, which can be formed by real humans~\cite{He2024}, or AI bots~\cite{Hintze2026}.

\section{Achieving change by promoting awareness of behavior}


As discussed in the section above, different methods have been considered for achieving collective change in both coordination games and public goods games, under the evolutionary game theory framework. Here, we build upon the classical quote of ``knowing is half the battle,'' and consider achieving collective change through interventions that focus on promoting awareness in the population. A unique aspect of focusing on awareness is that we can leverage social and behavioral processes that are more viable as intervention strategies for policymakers. For instance, the committed minority approach (including with targeted placement) can be ethically challenging to implement since it involves direct control over some players and the ability to manipulate the network structure. The awareness-based approaches presented here can reasonably be achieved by marketing and public message campaigns that can be designed with the help of subject matter experts. 

\begin{pullquote}
A unique aspect of focusing on awareness is that we can leverage social and behavioral processes that are more viable as intervention strategies for policymakers\end{pullquote}

A critical limitation of models that rely on solely the mechanism of coordination games is that they are only capable of capturing social influence and social conformity as key drivers of the emergence of collective behaviors. We focus on two particular processes recently identified in the social psychology literature as being potentially effective interventions for achieving change in an individual's behavior. These are an individual's sensitivity and reaction to observing emerging and new trends at the population level, and increasing visibility of a minority behavior. In this section, we show how these processes can be modeled within the coordination games formalism, and demonstrate that these drivers increase awareness across the population and, as a consequence, reduces the size of the committed minority needed to ultimately achieve collective change.



Awareness in public goods games is treated in the sequel. A key reason is that promoting awareness of behavior can be effective for scenarios described by coordination games, where social coordination and conformity are key. On the contrary,  for public goods games, making someone more aware that others are cooperating is unlikely to change their strategy due to the fundamental dilemma of defecting (maximizing individual benefit) and cooperating (maximizing the collective good). This is especially apparent because, as elaborated in Example~\ref{ex:coordination}, coordination games on networks can have multiple Nash equilibria whereas linear-reward public goods games have a unique Nash equilibrium at the all-defection profile. Hence, under evolutionary best-response updating, a population whose players engage in a public goods game will always converge to the unique Nash equilibrium, and raising awareness (following the approaches we present in this section) will not impact such an outcome. In the sequel, we present methods that go beyond raising awareness of the behavior of others, by considering opinions of preferred strategies, illustrating how these approaches can be leveraged to achieve collective change and favor cooperation in public goods games.


\subsection{Awareness via sensitivity to emerging trends}

The first intervention is inspired by recent literature on ``dynamic norms.'' Classical theory on norms suggests that a person's behavior can be substantially influenced by what they perceive to be normative in their surrounding (for instance, what is the behavior that the majority are adopting, what behavior is approved by the majority)~\cite{bicchieri2005grammar}. This is aligned with the concept of coordination games, in which players adopt the most popular strategy among their neighbors. Dynamic norms differs from this theory by demonstrating that an individual can be nudged to adopt a minority behavior when the individual has been made aware that said behavior has become more popular. In other words, people become sensitive to an emerging trend in the population. 

Researchers have applied dynamic norms to change behavior in many different contexts, including women's participation in STEM degrees and careers~\cite{cheng2020join}, the adoption of sustainable behaviors~\cite{sparkman2017dynamicnorm_sustainable}, and even participation in collective action~\cite{cohen2026novel}. This literature can also be linked to earlier studies on the ``spiral of silence,''~\cite{noelle1993spiral} and fads and fashions~\cite{bikhchandani1992informational_cascade}. In the context of social diffusion, an online experiment demonstrated that sensitivity to trends is not only present in the individual-level decision-making mechanisms, but it can also critically shape the emergent diffusion dynamics~\cite{ye2021nat}. Numerical simulations performed on the experimentally-validated model suggested that sensitivity to trends can unlock collective change~\cite{ye2021nat}. Building on this, we illustrate how it is possible to capture the sensitivity to emerging trends within the evolutionary game-theoretic framework for decision-making in \eqref{eq:coord_BR}. Then, we use the obtained model to study how leveraging this sensitivity create awareness to promote collective change, including in scenarios where no change would have otherwise occurred.

Mathematically, an individual $i$ who revises their strategy under the influence of emerging trends adopts the strategy that is becoming more popular. Using the definition of fraction of adopters of action $+1$ (or innovators) from \eqref{eq:zeta_t}, then a trend-seeking individual revises their strategy according to
 \begin{equation}\label{eq:trend}
        x_i(t+1)=\left\{\begin{array}{ll}+1&\text{if }\zeta(t)>\zeta(t-1),\\
        -1&\text{if }\zeta(t)<\zeta(t-1),\\
        x_i(t)&\text{if }\zeta(t)=\zeta(t-1).\end{array}\right.
    \end{equation}

\begin{figure}
    \centering
 \subfloat[]{\input{fig/trend1}}\subfloat[]{\input{fig/trend2}}
    \caption{Mechanism of the network coordination game with sensitivity to emerging trends in \eqref{eq:decision}. In (a), according to probability $u_t$, some individuals are selected to follow the emerging trend (orange area); the others follow the coordination game (green area) and generate $k=3$ interactions each.  Nodes in cyan (red) adopt the innovation, $+1$ (status quo, $-1$). In (b), individuals in the green area revise their strategy according to a coordination game with $\alpha=0.5$, those in the orange area follow the trend (assuming $z(t)>z(t-1)$, they switch action to $+1$). Then, links and regions are removed and the dynamics resumes from the next iteration. }
    \label{fig:trend}
\end{figure}
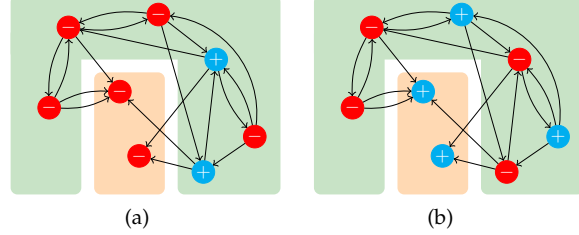

In the model originally proposed in~\cite{cdc2021}, and refined in~\cite{zino2022nexus}, this mechanism is combined with a coordination game to represent realistic decision-making processes. In particular, we introduce a parameter $u_t\in[0,1]$ that captures the \emph{sensitivity} of the population to emerging trends. At each time $t\in\mathbb N$, and independent of other individuals and past occurrences, individual $i$ revises their strategy as
\begin{equation}\label{eq:decision}
x_i(t+1)=\left\{\begin{array}{ll}
\eqref{eq:coord_BR}&\text{ with probability }1-u_t,\\
\eqref{eq:trend}&\text{ with probability }u_t.
\end{array}\right.
\end{equation}
In other words, $u_t$ quantifies the probability (sensitivity) that the individual makes their decision based on the perceived increasing popularity rather than a desire to coordinate, as illustrated in Fig.~\ref{fig:trend}.

In real-world settings, individuals often observe the behavior of a different set of players at each time step. For this reason, it is reasonable to assume that the network of social interactions changes in time, that is,  matrix $A$ is actually a time-varying matrix $A(t)$. This does not yield any fundamental change in the dynamics, besides introducing a dependence on $t$ in \eqref{eq:coord_BR} through the time-varying matrix $A$, which already depends on time throughout $\vect x(t)$.

In particular, the social psychology literature suggests that people are typically influenced by a bounded (and often fixed) amount of social contacts~\cite{Dunbar1992}. For this reason, we assume that at the generic time step $t$, $i\in\mathcal V$ observes the actions of a fixed number 
of \emph{social contacts} $k\in\mathbb N_+$, with $k\geq 2$, selected according to a mechanism inspired by discrete-time activity-driven networks~\cite{Perra2012adn}. Specifically, each social contact is selected uniformly at random from the entire population, independent of the others, with
\begin{equation}\label{eq:adn}
  \mathbb P[j\text{ is the $\ell$th contact of $i$ at time $t$}]=\frac1n.
\end{equation}
The neighbors of  $i$ at time $t$, denoted by $\mathcal N_i(t)$, are the individuals contacted by $i$ at that time. This generates a directed time-varying graph of interactions with $w_{ij}(t)=\frac{1}{k}$ if $j\in\mathcal N_i(t)$, 
which is then used in the action revision dynamics in \eqref{eq:decision}, in the case of \eqref{eq:coord_BR}. 

The randomness in the network formation process described by \eqref{eq:adn} and in the revision in \eqref{eq:decision} yield a stochastic process, which in general is not a Markov chain due to the dependence of \eqref{eq:trend} not the state at both $t$ and $t-1$. This makes analysis nontrivial, but still possible. In particular, through a probabilistic analysis of the dynamics that emerge from \eqref{eq:decision} based on the recursive use of Hoeffding's inequalities~\cite{Hoeffding1963}, it is possible to establish rigorous guarantees on the role of parameter $u_t$ in solving the collective change problem. First, it is worth noticing that in the absence of sensitivity to trends, that is, $u_t=0$,  collective change occurs with high probability if and only if one of the following two conditions are satisfied~\cite{zino2022nexus}: i) $\alpha>k-2$, or  $\zeta(0)>\zeta^*_{k,\alpha}$, where $\zeta^*_{k,\alpha}$ is the unique solution of $\Pi_{k,\alpha}(\zeta)=\zeta$ in $(0,1)$, with 
\begin{equation}\label{eq:def_pi}
\Pi_{k,\alpha}(\zeta):=\sum_{\ell=\left\lfloor k/(2+\alpha)\right\rfloor+1}^k\binom{k}{\ell}\zeta^\ell(1-\zeta)^{k-\ell}\,.
\end{equation}
In other words, collective change occurs if either the innovation is sufficiently superior to the status quo, or there are enough initial adopters of the innovation, and this is consistent with the literature discussed above. However, these conditions are quite restrictive. In the absence of relative advantage ($\alpha \approx 0$), $\zeta^*_{k,0} = 0.5$. That is, we require more than $50\%$ of the population initially adopting the innovation to guarantee collective change, which is not realistic in most real-world scenarios. Meanwhile, the first condition would require $\alpha > 1$ if each individual has just $k = 3$ network contacts, suggesting the innovation needs to offer twice the payoff of the status quo to guarantee collective change.

Instead, when individuals are sensitive to trends, the barriers for collective change are substantially lowered, as summarized in the following theorem from~\cite{zino2022nexus}.
\begin{theorem}[Effectiveness of trend-based  interventions]\label{th:gamma}
Consider the model in \eqref{eq:decision} with $k\geq 2$ and let $\zeta(0)=\frac{1}{2n}\sum (1+x_i(0))$ be the fraction of initial adopters. Then, collective change occurs with high probability if one of the following conditions is satisfied:
\begin{enumerate}
\item If $\alpha>k-2$, for any $\zeta(0)>0$;
\item If $\alpha<k-2$ and  $u_t>u^*_t$, for any $\zeta(0)>0$;
\item Otherwise, if $\zeta_0>\zeta^*_{k,\alpha,u_t}$,
\end{enumerate}
for the threshold values 
\begin{align}&u_t^*:=\inf\{u\in[0,1]:f_u(\zeta)>0,\,\text{for all } z\in(0,1)\}\,,\label{eq:gamma_star}\\
&\zeta^*_{k,\alpha,u_t}:=\inf\{\zeta\in[0,1]:f_u(\zeta)>0,\,\text{for all } z\in( \zeta,1)\} \label{eq:beta_star}\,,
\end{align}
where $f_{u_t}(\zeta):=(1-u)\Pi_{k,\alpha}(\zeta)-\zeta+u_t$.
\end{theorem}

\begin{figure}
    \centering
    \subfloat[]{\input{fig/nexus1}\label{fig:nexus_a}}
    \subfloat[]{\input{fig/nexus2}\label{fig:nexus_b}}
    \caption{Occurrence of collective change (with high probability) in coordination games when promoting awareness via sensitivity to emerging trends, with $k=3$. In (a), collective change occurs in the cyan region; in the gray region, it occurs if there are enough initial adopters. In (b), with $\alpha=0$, collective change occurs or does not occur if the fraction on initial adopters is in the cyan or red region, respectively. }
    \label{fig:nexus}
\end{figure}
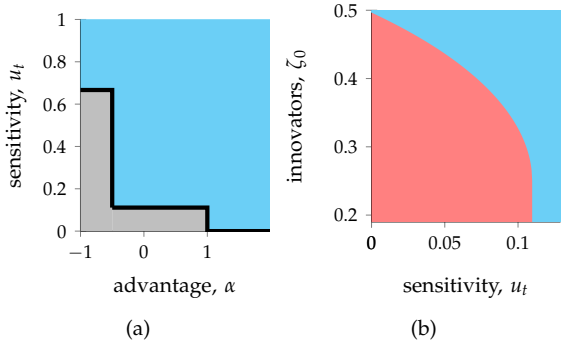

The first condition indicates that collective change always occurs if the innovation is sufficiently superior. Besides this, Theorem~\ref{th:gamma} establishes that creating awareness by increasing trend-based interventions can facilitate collective change in two main ways. First, if the sensitivity $u_t$ is sufficiently large (precisely, greater than a threshold $u^*_t$, that depends on $\alpha$ and $k$), then collective change occurs for any fraction of initial adopters. Second, even if $u_t < u^*_t$, sensitivity to trends can reduce the fraction of initial adopters needed to guarantee collective change, since the threshold $\zeta^*_{k,\alpha,u_t}$ is monotonically non-increasing in $u_t$. The impact of sensitivity-based interventions can be observed in Fig.~\ref{fig:nexus}, which is obtained with $k=3$. In Fig.~\ref{fig:nexus_a}, the cyan area illustrates where collective change is obtained for any initial fraction of adopters: if sensitivity is sufficiently large, then collective change can always be guaranteed. In Fig.~\ref{fig:nexus_b}, we consider the case $\alpha=0$, illustrating how the threshold in the initial condition $\zeta^*_{k,\alpha,u_t}$ is monotonically decreasing in the sensitivity to trends $u_t$. Above $\zeta^*_{k,\alpha,u_t}$, collective change is guaranteed with high probability. 

In summary, the theoretical derivations from~\cite{cdc2021,zino2022nexus} provide insights into how sensitivity to emerging trends can be leverage to unlock collective change when the relative advantage and the size of the committed minority are not sufficient to trigger an adoption cascade based on solely social coordination and peer pressure. Sensitivity to emerging trends can create awareness of the ongoing change and on its increasing popularity, guiding the population to a collective change.

\subsection{Enhancing visibility to create awareness}

The next intervention begins with the observation that an individual's willingness to adopt a behavior is strongly impacted by the visibility of others adopting said behavior, as this makes it more normative and salient~\cite{bicchieri2005grammar}. In the context of promoting transition towards sustainability, several empirical studies have examined the use of stickers or other material to signal a person's adherence to some given pro-environmental behavior that is currently not widely adopted~\cite{HamannEtAl2015,reese2013sticker,merkelbach2021committing}. The use of such visible signals helped  increase others' awareness, and nudge non-adopters to follow suit.


Building on the model described in \eqref{eq:decision}, we can further expand it to incorporate visibility-based interventions. We assume that visibility affects the network formation process, whereby people who adopt the innovation are more visible than those who do not, and thus an individual has a higher probability of observing them. Inspired by activity-driven networks with attractiveness~\cite{Alessandretti2017}, we introduce a parameter $u_v\geq 0$ that represents the relative increase in visibility of innovation adopters. We then replace \eqref{eq:adn} with
\begin{equation}\label{eq:aadn}\begin{array}{l}
 \mathbb P[j\text{ is the $\ell$th contact of $i$ at time $t$}]=\\
\qquad\qquad\qquad\qquad \left\{\begin{array}{ll}
    \frac{1+u_v}{n(1+\zeta(t))}&\text{ if }x_j(t)=1,\\   \frac{1}{n(1+\zeta(t))}&\text{ if }x_j(t)=0,
    \end{array}\right.\end{array}
\end{equation}
which accounts for the increased probability of interacting with adopters of the innovation. In is worth noticing that \eqref{eq:aadn} yields an adjacency matrix $A(t)$ that is not only time-varying and subject to stochasticity, but also with a state-dependent probability distribution law, since the presence of an edge depends on the value of $\zeta(t)$. This increases the complexity of the dynamics and complicates the analysis. However, also in this setting, probability-based tools can be used to establish conditions under which collective change is achieved. All details can be found in~\cite{zino2023ifac}.

\begin{figure}
    \centering
\input{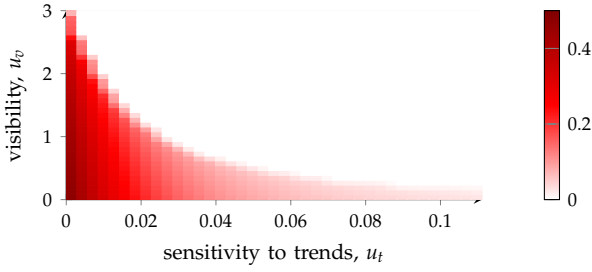} 
\caption{Awareness in coordination games via sensitivity to emerging trends and visibility. Threshold for the fraction of initial adopters that guarantees collective change for $k=3$ and $\alpha=0$, for different values of sensitivity to trends $u_t$ and increased visibility of innovators $u_v$. }
    \label{fig:visibility}
\end{figure}

Increasing visibility can meaningfully decrease the fraction of initial adopters $\zeta(0)$ needed to guarantee collective change. Namely, with reference to Theorem~\ref{th:gamma}, increasing the visibility of those adopting the innovation can decrease the threshold $\zeta^*_{k,\alpha,u_t}$ computed in the theorem. In particular, the threshold becomes a function of $u_t$, denoted as $\zeta^*_{k,\alpha,u_t,u_v}$, and it can be shown that this is a non-increasing function of $u_v$. For the rigorous extension of Theorem~\ref{th:gamma} that includes explicit derivations of such expressions, we refer to~\cite{zino2023ifac}. Here, we consider an illustrative scenario involving $k=3$ and  $\alpha=0$, with the threshold  $\zeta^*_{k,\alpha,u_t,u_v}$ shown in Fig.~\ref{fig:visibility} --- this can be examined in comparison to Fig.~\ref{fig:nexus}(b). Notice that the increasing sensitivity to trends, and also increasing visibility, can each separately reduce the critical fraction of initial adopters needed to guarantee collective change. Interestingly, Fig.~\ref{fig:visibility} suggests that effective intervention policies can be designed by combining both types of actions (that is, simultaneously increasing the visibility of adopters and enhancing sensitivity to trends) in order to create awareness of the ongoing collective change and, ultimately, ensuring its successful occurrence. This paves the way for research designing a combination of these two action in a cost-effective manner by formalizing an optimization problem using the exact analytical derivation of the thresholds from~\cite{zino2023ifac}.

In the last two sections we have illustrated how, building on the game-theoretic framework of coordination games, it is possible to incorporate further mechanisms that capture different types of interventions to generate awareness of collective change in a population by highlighting emergent trends or enhancing the visibility of those who adopt the innovation. There are two different perspectives to interpreting the impact of these mechanisms. First, we can consider sensitivity to trends and increased visibility (and in particular specific values of $u_t$ and $u_v$) as characteristics inherent to the scenario of interest, and the analytical results help to determine whether collective change happens or not. This may be relevant if we consider visible versus non-visible sustainable behavior (cycling is more visible than reducing energy consumption), or different social environments (online social media may favor trends compared to offline group discussions). Alternatively, we can consider a policymaker who is interested in implementing interventions based on increasing sensitivity to trends (for example, by providing information on such trends~\cite{
Loschelder2019,cheng2020join}) and visibility of the desired behavior (for instance, using stickers or other signals~\cite{reese2013sticker,merkelbach2021committing}) in order to achieve collective change in scenarios for which it would otherwise not occur. The proposed mathematical modeling framework means that for the latter, we have established  quantitative results on the effectiveness of such interventions and, ultimately, policymakers can leverage optimization techniques to design optimal intervention policies to favor collective change.

\begin{sidebar}{Opinion dynamics}
\section[Opinion dynamics]{}\label{side:opinion}

\setcounter{sequation}{3}
\renewcommand{\thesequation}{S\arabic{sequation}}
\setcounter{stable}{0}
\renewcommand{\thestable}{S\arabic{stable}}
\setcounter{sfigure}{0}
\renewcommand{\thesfigure}{S\arabic{sfigure}}

\sdbarinitial{W}e present a brief digression on mathematical models of opinion and belief formation due to social influence over networks. This approach emerged from French's seminal effort to understand social power~\cite{french1956_socialpower_SA}, and has its foundation in the seminal DeGroot~\cite{degroot1974OpinionDynamics_SA} and Friedkin--Johnsen~\cite{friedkin1990_FJsocialmodel_SA} models. 

\subsection{Opinions as continuous variables}
Consider a network of $n \geq 2$ individuals $\mathcal{V} = \{1,2,\dots,n\}$, discussing a given statement or issue over discrete-time steps $t =  0, 1, \hdots$. Individual~$i$ is characterized by the state variable $y_i(t) \in \mathbb R$, referred to as individual $i$'s opinion or belief, representing their cognitive orientation towards a statement or issue. It is reasonable to assume that opinions are bounded in a domain symmetric with respect to the origin. For example, we can assume that $y_i \in [-1,1]$, where  $y_i < 0$ and $y_i > 0$ represent disagreement or agreement with the specific statement or issue of discussion (and $y_i = 0$ a neutral position). 

\subsection{Linear averaging processes}
Individuals communicate their opinions over a weighted graph $\mathcal{G} = (\mathcal{V}, \mathcal E,  W)$, where $W$ is a stochastic matrix, that is, all of its entries are nonnegative and $\sum_{j\in\mathcal V}w_{ij}=1$ for all $i$. There holds $w_{ij} > 0$ if and only if $i$ is directly influenced by $j$'s opinion, and $w_{ij}$ represents the relative social influence weight that individual $j$ exerts on individual $i$ (relative because $\sum_{j\in\mathcal V}w_{ij}=1$).

At every time step, the opinion of individual~$i\in\mathcal{V}$ evolves according to
\begin{sequation}\label{eq:fj_individual}
    y_i(t+1) = \sum\nolimits_{j\in \mathcal V}w_{ij}y_j(t),
\end{sequation}
which describes updating via a convex combination of the current opinions of $i$'s neighbors. Further terms can be added to the dynamics to capture, for instance, existing prejudices or biases~\cite{friedkin1990_FJsocialmodel_SA}, negative social influence~\cite{altafini2013antagonistic_interactions_SA}, or bounded confidence~\cite{bernardo2024_SA}. For more details, we shall refer to~\cite{Flache_SA,Noorazaz_SA}.

\subsection{Emergent behavior}

The emergent behavior of the opinion dynamics in \eqref{eq:fj_individual} has been extensively studied. In particular, opinions always remain bounded between the maximal and minimal initial opinions and, under mild conditions on the connectivity  of the network (specifically, if the network has a globally reachable node), they converge to a consensus $y(t)\to \bar y {\bf 1}$. The final consensus value $\bar y$ is shaped by the structure of the network, and is in fact a weighted average of the initial conditions, $\bar y=\pi^\top y(0)$, as illustrated in Fig.~\ref{fig:consensus}. The weight vector $\pi$ is in fact the normalized left eigenvector associated with the unit eigenvalue of $W$, with entries positive and summing to $1$. Entry $\pi_i$ thus describes the social power of node~$i$ in shaping the discussion outcome. 
Seminal works have explored further features of the emergent behavior of opinion dynamics, including the rate of convergence, and the emergence of more complex phenomena in the scenario of the presence of nonlinearities in \eqref{eq:fj_individual}. An excellent survey is found in \cite{proskurnikov2017tutorial_SA}.

\sdbarfig{\input{fig/consensus}}{Linear averaging process. Trajectories of the linear averaging process in \eqref{eq:fj_individual} for $6$ individuals, with initial opinions selected uniformly at random in $[-1,1]$. The plot shows convergence to a consensus on a weighted average of the initial opinions. \label{fig:consensus}}

\subsection{Opinion dynamics as games}

Interestingly, the opinion dynamics in \eqref{eq:fj_individual} can be equivalently cast in a game-theoretic framework~\cite{Marden2009_SA,Ghaderi2014opinion_SA}. In fact, by defining the payoff function
\begin{sequation}\label{eq:opinion_game}
    f_i(s,\vect y_{-i})=-\frac12\sum\nolimits_{j\in\mathcal V}w_{ij}(s-y_j)^2,
\end{sequation}
the dynamics of \eqref{eq:fj_individual} can be interpreted as a best-response update rule, that is, $y_i(t+1)=  \mathfrak{B}_i (f_i(\cdot, \vect y_{-i}(t)))$. This parallelism between opinion dynamics and games is advantageous for the integration of opinion dynamics in game-theoretic models of decision-making, and for theoretical analysis that borrows tools from evolutionary game theory.

\end{sidebar}

\section{Opinions matter}

So far, we have discussed how different behavioral mechanisms such as sensitivity to trends and the impact of increased visibility can create awareness that favors collective change. Then, we illustrated how these mechanisms can be readily encapsulated within an evolutionary game-theoretic framework for decision-making processes, offering an analytically-tractable model to study how these mechanisms can be leveraged as awareness-based interventions to promote collective change.

However, besides social coordination and the mechanisms already presented, there is abundant evidence in the social and behavioral sciences literature on how an individual's opinions play a critical role in their decision-making processes. 
While it is true that opinions and actions are often highly correlated (see for instance the seminal `theory of planned behavior'~\cite{ajzen1991theory}), we point out that there is particular interest in the opposite. That is, the observation that often the actions and opinions of (many) people are not aligned. A classic example is the phenomenon of unpopular norms~\cite{willer2009false_enforcement,Smerdon2019}, in which a community keeps adopting a collective behavior that is disapproved by the majority. Pluralistic ignorance is a conceptually similar phenomenon, in which people incorrectly assume there is widespread support for the status quo and limited approval of the innovation. This is the case for several instances of sustainable behaviors~\cite{Geiger2016}, and has been observed in a variety of contexts~\cite{prentice1993pluralistic,ogorman1975pluralistic}. These two phenomena have obvious specific relevance to collective change. 

On the one hand, opinions can play a critical role in hindering collective change. 
For instance, despite there being unequivocal evidence that Earth is warming at an unprecedented rate and human activity is the principal cause~\cite{nasa}, political polarization and a divergence of beliefs on the adoption of sustainable behavior has hindered collective change~\cite{Judge2023}. On the other hand, our key idea is that opinion sharing can instead lead to public opinion shifting to support collective change, which increases awareness that ultimately leads to the changes in behavior we wish to observe. Indeed, sharing opinions may be especially valuable, nay critical, in the presence of pluralistic ignorance or unpopular norms, to break out of the status quo and achieve collective change. 



\begin{pullquote}
    Opinion sharing can be used to create awareness and ultimately lead to collective change
\end{pullquote}

We argue that there is a key distinction between opinion formation and decision-making (although they are coupled and influence one another), since the former is related to a person's beliefs and attitudes, while the latter is related to their actions and behaviors. Thus, although opinion dynamics have been extensively studied in the systems and control literature as discussed in ``\nameref{side:opinion}," opinion dynamics models have been typically treated independently from decision-making models or the two problems are conflated as being the same. This limits the ability to study the impact of opinion formation processes on human behavior and investigating how collective change can be achieved by intervening on such a process to generate awareness. There is a need to develop new mathematical frameworks that encapsulate opinion formation  into collective decision-making dynamics, which is our focus for the remainder of this article.

\subsection{Classical approaches to account for opinions}

Before introducing the coevolutionary framework for actions and opinions, we revisit some classical models that embed opinion dynamics with decision-making dynamics. In~\cite{Martins2008coda}, the
continuous-opinion discrete-action (CODA) model was proposed, which along with its extensions~\cite{Chowdhury2016coda,Ceragioli2018quantized,Tang2021coda} has been used to study innovation diffusion~\cite{Martins2009innovation} and competition in duopolies~\cite{Varma2017coda}. The underlying assumption of the CODA model is that individuals have (private) opinions and (public) actions. An individual revises their opinion by observing the actions of others (according to a classical opinion dynamics model), and then make their decision on which action to take as a direct quantization of their private opinion, or based on a probabilistic mechanism~\cite{Ravazzi2026}. Another related framework is that of Expressed and Public Opinion (EPO) models~\cite{duggins2017psychologically,ye2019influence,cheng2022social,Shang2021_epo,Banisch2018,gaisbauer2020dynamics,kaminska2025impact}. While different models have been proposed, they all consider the coevolution of private and expressed opinions, whereby a person does not always express their true (private) opinion, but their private opinion is influenced by what is expressed by others. However, no decision-making process is incorporated. 

On the other hand, decision-making models have been proposed which take into account an individual's private opinion, which can be constant~\cite{centola2005emperor,Jadbabaie2023}, 
or time-varying~\cite{Park2021}. A body of literature has developed a framework for the coevolution of behaviors and beliefs~\cite{gavrilets2026cultural_pluralistic_ignorance,Gavrilets2024_norm_belief,gavrilets2024coevolution_dilemma_beliefs}, sharing some similarities but also key differences with the framework presented in this article. Namely, individuals update their behavior by best-responding to a utility function that comprises several terms capturing different games (such as a public goods game), while opinions evolve according to continuous-time flows that balance a desire to act according to one's belief and a pressure to conform to the observed normative behavior. This allows for rich emergent behavior, but the model complexity limits analytical tractability.

In summary, existing models capture some aspects of realistic opinion sharing and their impact on decision-making, but almost all lack the ability to capture nontrivial interactions between actions and opinions, or have limited analytical tractability. 
In the next section, we introduce a coevolutionary model framework that unifies opinion dynamics and collective decision-making, allowing for a natural approach to describing the evolution of opinions and actions for different problems of interest. Critically, the framework is analytically tractable due to its development within the evolutionary game theory paradigm, allowing for the derivation of convergence guarantees, characterization of the asymptotic behavior, as well as extensions that consider problems of control and influence.

\section{Coevolutionary framework for opinions and behavior}

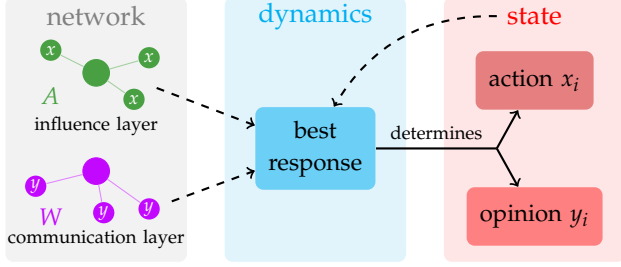
\begin{figure}
    \centering
    \input{fig/schema}
    \caption{Schematic of the coevolutionary framework. }
    \label{fig:schematic}
\end{figure}

To address the limitations of the approaches described in the previous section and provide a mathematical framework in which it is possible to explicitly leverage opinion dynamics to create awareness and guide human decision-making, we now present a coevolutionary framework for actions and opinions. This framework, originally proposed in~\cite{zino2020chaos,Zino2020cdc}, and then formalized in an evolutionary game-theoretic fashion in~\cite{Hassan2023tac}, is illustrated in Fig.~\ref{fig:schematic}. 

Individuals $\mathcal V = \{1,\hdots,n\}$ are characterized by a two-dimensional state variable $z_i(t)=[x_i(t),y_i(t)]$ evolving over discrete times $t\in\mathbb N$. Individual $i\in\mathcal V$ adopts a binary action $x_i(t)\in\{-1,+1\}$ and holds an opinion $y_i(t)\in[-1,+1]$ on their support for the action: $y_i(t)=-1$ and  $y_i(t)=+1$ represent individual~$i$ having maximal support for action $-1$ and $+1$, respectively. Actions and opinions are gathered in vectors $\vect{x}(t)\in\{-1,1\}^n$ and  $\vect{y}(t)\in[-1,1]^n$, and the state of the system is the joint $2n$-dimensional vector $\vect{z}(t):=(\vect{x}(t),\vect{y}(t))\in\{-1,1\}^n\times [-1,1]^n$. We use the standard game-theoretic notation $\vect{z_{-i}}$ to denote the state vector without the $i$th entries.

Individuals simultaneously revise their action and opinions, according to an evolutionary game-theoretic mechanism. In particular, individuals interact on a two-layer network, whereby they can observe each others actions on an influence layer $(\mathcal V,\mathcal E_A,A)$ and share their opinions on a communication layer $(\mathcal V,\mathcal E_W,W)$. Hence, the entries $a_{ij}$ and $w_{ij}$ of the two (weighted) adjacency matrices $A$ and $W$ capture the influence of $j$'s action and $j$'s opinion on $i$'s dynamics, respectively.

The payoff function is defined to capture three key mechanisms: i) the game that governs the individual's action (for example, a coordination game or a public goods game); ii) opinion formation through social influence, and iii) an individual's desire to have consistency between their action and opinion. By denoting the payoff function for a ``classical" decision-making game $\hat f_i(x_i,\vect x_{-i})$, and exploiting the game-theoretic formulation of opinion dynamics models as detailed in ``\nameref{side:opinion}," we propose a unified payoff function
\begin{align}
    f_i(\vect{z_i},\vect{z_{-i}})=&\displaystyle\gamma_i\hat f_i(x_i,\vect x_{-i})-\frac{1}{2}\beta_i\sum\nolimits_{j \in \mathcal V}{w}_{ij}(y_i-y_j)^2\nonumber\\&\displaystyle
   -\frac12 \lambda_i(x_i-y_i)^2,\label{eq:general_coevolutionary_payoff}
\end{align}
where $\gamma_i,\beta_i,\lambda_i \in \mathbb R_+$ represent~$i$'s relative weight of actions, opinions, and desire for consistency between these two aspects, respectively. Note 
the consistency term is the negative of the quadratic difference between $x_i$ and $y_i$, indicating that their payoff decreases when their opinion and action are not aligned.

Following the standard best-response dynamics (presented in ``\nameref{side:egt}"), a generic individual $i\in\mathcal R(t)$ who revises their strategy at time $t$ updates their actions and opinions simultaneously according to \eqref{eq:BR_generic}, with the convention that, when the best response comprises multiple elements, we set $x_i(t+1)=x_i(t)$. In other words, each individual $i\in\mathcal R(t)$ performs a joint best-response with respect to \eqref{eq:payoff_coordination}, as illustrated in Fig.~\ref{fig:schematic}. For the sake of readability, we report all the variables and parameters of the coevolutionary framework in Table~\ref{tab:coevo}.

\begin{table}
\center
\caption{Symbols for the coevolutionary framework.\label{tab:coevo}}
\begin{tabular}{|c|c|}
\hline
Symbol& Meaning \\
\hline
$x_i(t)$& Action of player $i\in\mathcal V$ at time $t\in\mathbb N$\\
$y_i(t)$& Opinion of player $i\in\mathcal V$ at time $t\in\mathbb N$\\
$z_i(t)$& State of player $i\in\mathcal V$ at time $t\in\mathbb N$\\
$\mathcal R(t)$& Players who revise state at time $t\in\mathbb N$\\
$\hat f_i$& Payoff function of the action game\\
$\gamma_i$& Impact of others' action on the decision making\\
$\beta_i$& Impact of others' opinion on the decision making\\
$\lambda_i$& Impact of self-consistency on the decision making\\
$a_{ij}$& Peer pressure due to $j$'action on $i\in\mathcal V$\\
$w_{ij}$& Social influence of $j$'opinion on $i\in\mathcal V$\\
\hline
\end{tabular}
\end{table}

\begin{figure*}
    \centering
    \subfloat[Complete graph ($\gamma_i=1$)]{\input{fig/complete.tex}\label{fig:reduce_th}}\quad
    \subfloat[Real-world network]{\input{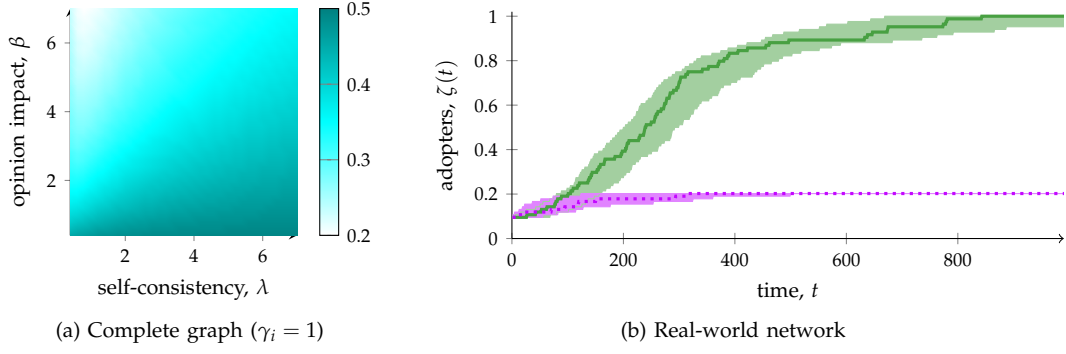}\label{fig:malawi}}
    \caption{Opinion sharing to facilitate collective change in coordination games, with $\gamma_i=1$, $\lambda_i=\lambda$, and $\beta_i=\beta$, for all $i\in\mathcal V$. In (a), the heatmap shows the critical mass needed to guarantee collective change according to the model in \eqref{eq:coevolutionary_payoff} for a complete graph; these results should be compared to a pure coordination game, where the critical mass is $50\%$.  In (b), we compare two representative trajectories of the coevolutionary model (solid green curve) and of a pure coordination game (dotted violet curve) on a network reconstructed from real-world interactions~\cite{malawi_net}, with the $8$ nodes with largest eigenvalue centrality selected to be committed minority. The colored areas represent the 95\% simulations envelope over 100 independent realizations of the processes. }
\end{figure*}

\section{Opinion Awareness for Collective Change}

We discussed above how an individual's opinion may have a critical impact in shaping actions, and thus on collective change. In the following, we discuss how the coevolutionary framework in \eqref{eq:general_coevolutionary_payoff} can be leveraged to explore interventions that focus on creating awareness through opinion sharing to achieve collective change. 

\subsection{Opinion sharing to favor collective change in coordination games}

Integrating \eqref{eq:payoff_coordination} into \eqref{eq:general_coevolutionary_payoff} leads to the payoff function
\begin{align}
    f_i(\vect{z_i},\vect{z_{-i}})=&\displaystyle\frac{\gamma_i}{4}\sum_{j \in \mathcal V} a_{ij} \Big( (1-x_j)(1-x_i) + (1+x_j)(1+x_i)\Big)\nonumber\\&\displaystyle -\frac12\beta_i\sum\nolimits_{j \in \mathcal V}{w}_{ij}(y_i-y_j)^2-\hspace{-.1cm}\frac12 \lambda_i(x_i-y_i)^2,\label{eq:coevolutionary_payoff}
\end{align}
 where, for the sake of simplicity, we set $\alpha=0$ (a nonzero relative advantage can be readily introduced).  

First, \eqref{eq:coevolutionary_payoff} enables us to derive a closed-form expression for the best-response update rule in \eqref{eq:BR_generic} with the payoff function in \eqref{eq:coevolutionary_payoff}. Following~\cite[Proposition~1]{Hassan2023tac}, individual $i\in\mathcal R(t)$ that updates their state does so as
\begin{subequations}\label{eq:dinamics}
\begin{align}
\label{x-dinamic}
&x_i(t+1) = s(\vect z(t)),\\
\label{y-dinamic}
&y_i(t+1)=\frac{\beta_i}{\beta_i+\lambda_i}\sum\nolimits_{j \in \mathcal V} {w}_{ij}y_j(t) + \frac{\lambda_i}{\beta_i+\lambda_i} s(\vect{z}(t)),
\end{align}
\end{subequations}
where 
\begin{equation}\label{eq:s}
    s(\vect z(t))=\begin{cases}
    +1 \qquad &\text{if}\; \delta_i(\vect z(t)) >0, \\
    -1 \qquad &\text{if}\; \delta_i(\vect z(t)) <0, \\
    x_i(t) \qquad &\text{if}\; \delta_i(\vect z(t)) =0,
\end{cases}
\end{equation}
with
\begin{equation}\label{eq:delta}
    \delta_i(\vect z(t)) =\gamma_i\sum\nolimits_{j\in\mathcal V}a_{ij}x_{j}(t)+\frac{2\beta_i\lambda_i}{\beta_i+\lambda_i}\sum\nolimits_{j\in\mathcal V}w_{ij}y_{j}(t),
\end{equation}
which has an easy-to-interpret structure. In fact, \eqref{x-dinamic} and \eqref{eq:s} prescribe that individuals decide which action to adopt (either $-1$ or $+1$), depending on a combination of peer pressure and social influence from others' opinions, regulated by the model parameters as per \eqref{eq:delta}. Then, they revise their opinion according to \eqref{y-dinamic}, which is a convex combination of the opinions of others (social influence à la opinion dynamics) and their own action (self-consistency).

By  analyzing \eqref{eq:dinamics}, several theoretical results have been established, including convergence to an equilibrium for all initial conditions by leveraging potential game theory~\cite{monderer1996potential}, and characterization of  equilibria of interest such as consensus and polarized configurations~\cite{Hassan2023tac,Liang2025}.

Within the focus of this article, we are interested in the possibility of leveraging the opinion formation process to generate awareness and favor collective change. To this aim, we consider the following scenario. At time $t=0$, the population is at the consensus equilibrium $\vect{x}(0)=\vect{y}(0)=-\vect{1}$, and we want to steer it to $\vect{x}=+\vect{1}$ (namely, achieve collective change) by acting on the opinions through a committed minority that steadily supports the desired opinion. We implement this by assuming that there is a target set $\mathcal C\subseteq \mathcal V$, such that we can set $x_i(t)=y_i(t)=+1$ for all $t\geq 1$ and $i \in \mathcal C$. In other words, we assume that the members of the committed minority not only adopts the innovation, but they also share opinions that are maximally supportive of it.

This problem has been formulated and studied in~\cite{raineri2024,raineri2025_tcns}. The fundamental technical result that enables the study of this problem is the proof that in the scenario described in the above, all trajectories $\vect z(t)$ of the dynamics are not only convergent, but also monotonically non-decreasing, as demonstrated in~\cite[Theorem~2]{raineri2025_tcns}. Building on this result, an algorithm was proposed to check in a computationally efficient manner whether introducing the committed minority in a target set is sufficient to guarantee collective change. 
Building on this algorithm, an effective heuristics to determine the minimal control set (that is, the minimum number of nodes needed to be controlled) was proposed. Interestingly, the algorithm can be applied analytically to network topologies with certain regularity and symmetry (for example, complete or star graphs) and numerically in an efficient manner on more complex and realistic network structures, providing insight into how generating awareness via opinion sharing can facilitate collective change. In fact, for a complete graph, in a standard coordination game with homogeneous parameters and no relative advantage ($\alpha=0$), \eqref{eq:coord_BR} implies that the target set should include $50\%$ of the nodes to achieve collective change. Awareness generated by means of opinion sharing can lower size of the target set, especially when opinions have a large impact on individuals decision making (when $\beta_i$ is large). This is illustrated in Fig.~\ref{fig:reduce_th}, where we report the critical mass of the committed minority needed to guarantee collective change as a function of the model parameters (placement in the network is irrelevant due to the symmetry of the network structure). Explicit computations can be found in~\cite[Proposition~3]{raineri2024}. More realistically, Fig.~\ref{fig:malawi} compares the trajectory of the coevolutionary model (in blue) and of a pure coordination game (in red) with the same target set on a realistic network, reconstructed from real-world interaction data on a village in rural Malawi~\cite{malawi_net}. These insights reinforce the analytical insights gained from the complete network: Awareness generated via opinion sharing can be a key driver to facilitate collective change. 

\subsection{Promoting cooperation in social dilemmas}

\begin{figure*}
    \centering
    \begin{minipage}{0.25\linewidth}
        \subfloat[Initial actions]{\includegraphics[width=\columnwidth]{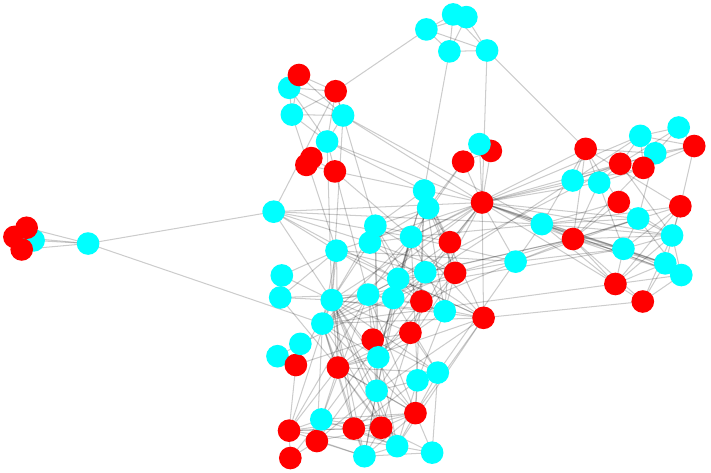}\label{fig:init_pgg}}
    \end{minipage}
    \begin{minipage}{0.45\linewidth}
        \subfloat[Opinions without committed minority]{\includegraphics[width = \columnwidth]{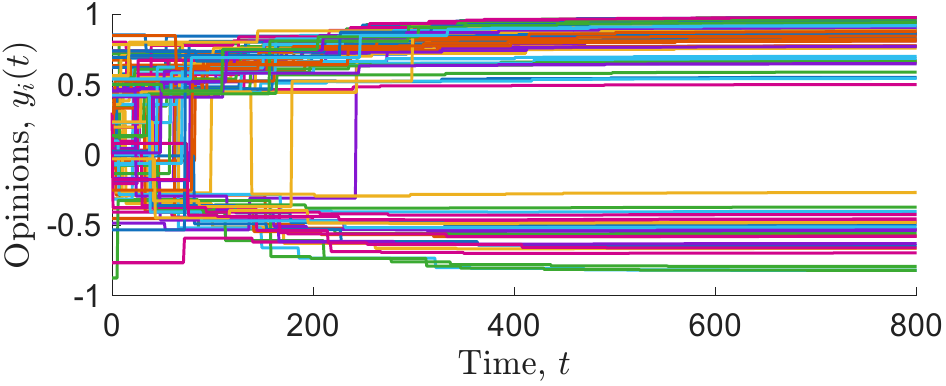}\label{fig:pgg_opinions_nochange}}\vfill
    \subfloat[Opinions with committed minority]{\includegraphics[width =\columnwidth]{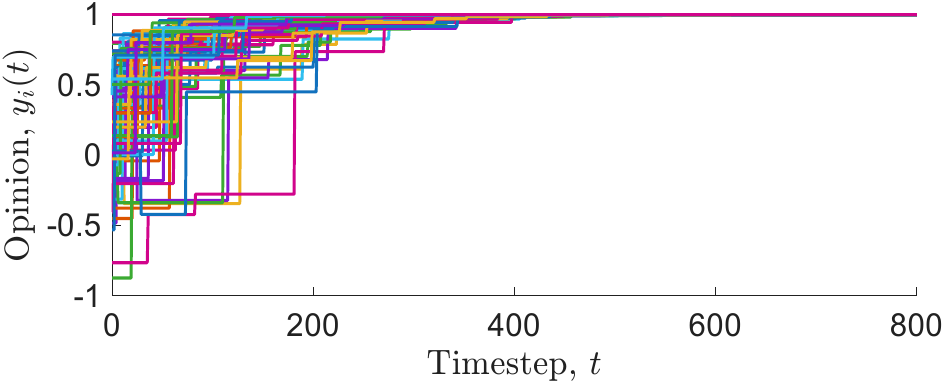}\label{fig:pgg_opinions_change}}
    \end{minipage}
    \begin{minipage}{0.25\linewidth}
        \subfloat[Final actions without committed minority]{\includegraphics[width = \columnwidth]{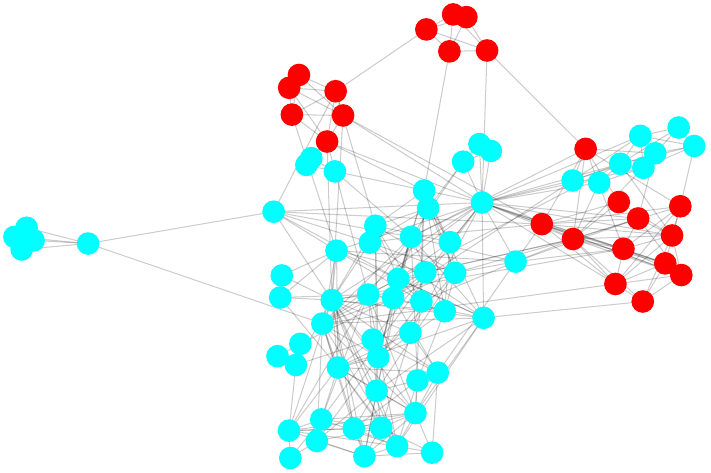}\label{fig:pgg_actions_nochange}}\vfill
    \subfloat[Final actions with committed minority]{\includegraphics[width = \columnwidth]{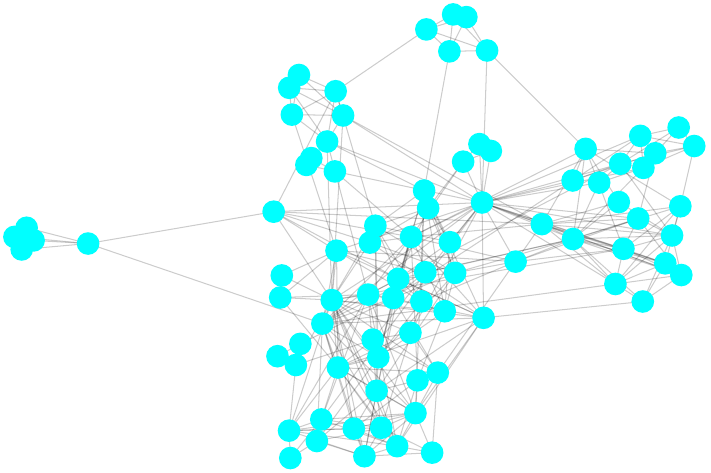}\label{fig:pgg_actions_change}}
    \end{minipage}
    \caption{Impact of sharing opinion in public goods games. Panel (a) shows the initial actions with defectors and cooperators in red and cyan, respectively. Panel (b) and (d) showcase the dynamics of the population when there are no committed minority, converging to a clustered equilibrium. Panel (c) and (e) showcase the dynamics when the 15  highest eigenvector centrality nodes are set as the committed minority, resulting in convergence to the all-cooperation equilibrium. \label{fig:pgg_change} }
\end{figure*}

The coevolutionary framework discussed in the above can also be tailored to a population faced with a social dilemma. In this case, by integrating the payoff function of the public goods game in \eqref{eq:payoff_pgg} into \eqref{eq:general_coevolutionary_payoff}, we obtain
\begin{equation}\label{eq:coevolutionary_payoff_pgg}
{{
\begin{array}{l}
    f_i(\vect{z_i},\vect{z_{-i}})=\displaystyle\gamma_ir\Big(\frac{n-1+\sum_{j\in\mathcal V\setminus\{i\}} x_j}{2n}\Big)+\gamma_i\Big(\frac{r}{n}-1\Big)\frac{x_i+1}{2}\\\qquad\qquad\quad\displaystyle-\frac12\beta_i (1-\lambda_i)\sum\nolimits_{j \in \mathcal V}{w}_{ij}(y_i-y_j)^2\hspace{-.1cm} -\hspace{-.1cm}\frac12 \lambda_i(x_i-y_i)^2.
\end{array}}}
\end{equation}

Just as with the scenario where the actions are described by a network coordination game, it is possible to derive a closed-form expression for the update rule for a generic individual $i\in\mathcal R(t)$ best-responding to \eqref{eq:coevolutionary_payoff_pgg}. In particular,~\cite[Proposition 1]{connor_ecc} yields exactly the same update rule in \eqref{eq:dinamics}, with however a different discriminant function
\begin{equation}\label{eq:delta2}
    \delta_i(\vect z(t)) =  \gamma_i\left( \dfrac{r}{n}-1 \right) + \frac{2\beta_i\lambda_i}{\beta_i+\lambda_i}\sum\nolimits_{j\in\mathcal{V}}w_{ij}y_j(t),
\end{equation}
that replaces \eqref{eq:delta}. Just as before, the dynamics have an intuitive interpretation, whereby individuals revise their action based on the discriminant in \eqref{eq:delta2}, and their opinion updates as as a weighted average of the opinions of others and of their own action. Interestingly, and different from the case of coordination games, \eqref{eq:delta2} depends on the state of the system only through the opinions $\vect y$, suggesting that for social dilemmas, opinions can have an even a stronger impact in shaping collective change. 

Consistent with the discussion on public goods games in Example~\ref{ex:pgg}, there is a Nash equilibrium in which all players defect (in fact, at this equilibrium, $\vect x = \vect y = -\vect 1$). However, the introduction of the opinion dynamics in shaping decision-making can result in other Nash equilibria for the coevolutionary model. For instance, analysis of the dynamics~\cite[Theorem~1]{connor_ecc} reveals there are conditions for the existence of an equilibrium in which all players cooperate, with $\vect x = \vect y = +\vect 1$. Namely, the following inequality is required to be satisfied for all $i\in\mathcal V$:
\begin{equation}\label{eq:pgg_cooperate_equib_cond}
    \frac{\beta_i\lambda_i}{\beta_i+\lambda_i}>\gamma_i\Big(1-\frac{r}{n}\Big).
\end{equation}

There has been limited analysis of the coevolutionary model in the context of public goods games, and much work remains to be pursued.  Here, we conclude this section by showcasing two exemplar simulations that reveal both the complex equilibria that can arise when we explicitly model the impact of opinion dynamics on people's decision-making, and the promise that committed minority can help to facilitate collective change even in the challenging case of social dilemmas. Figure~\ref{fig:pgg_change} presents two simulations on the real-world Malawi village network with the same initial  opinions, shown in Fig.~\ref{fig:init_pgg}, but different emergent collective phenomena. 
First, we simulate the coevolutionary model and we report the evolution of the opinions  and the final actions in Fig.~\ref{fig:pgg_opinions_nochange} and Fig.~\ref{fig:pgg_actions_nochange}, respectively. The network converges to an equilibrium with clusters of defectors and cooperators with opinions split into two camps, with one camp favoring cooperation and the other camp supporting defection. Already, we see how opinion formation processes can fundamentally change the long-term dynamics ---as discussed in Example~\ref{ex:pgg}, best-response updating will always result in all players defecting in the long term (that is the only Nash equilibrium of the game, even though not socially optimal). Here, we see that cooperation can persist due to the combined effect of people sharing opinions, and each person's desire for consistency between their action and opinion. The former allows a large part of the network to hold opinions in favor of cooperation, and the latter drives these individuals to cooperate despite the presence of others who defect.

Note that \eqref{eq:pgg_cooperate_equib_cond} is satisfied with the parameters of the simulation, so the all-cooperation configuration $\vect x = \vect y = +\vect 1$ is an equilibrium, but convergence does not occur to it. Next, we select the 15 nodes with highest eigenvector centrality to act as a  committed minority who support cooperation. Noting that the initial opinions and actions are the same for the non-committed people, and under the same activation sequence, the opinion dynamics are shown in Fig.~\ref{fig:pgg_opinions_change} and the final actions  in Fig.~\ref{fig:pgg_actions_change}. Here, the network converges to the all-cooperation  equilibrium, illustrating how a committed minority can facilitate collective change, achieving the social optimum. 

Finally, it is worth commenting that the committed minority would have no impact if they were only stubbornly selecting action $x_i = +1$, without acting as opinion leaders, since \eqref{eq:delta2} is independent of $\vect x$. 
Thus, collective change occurs precisely because the committed minority stubbornly hold and share opinions supporting cooperation, raising awareness and ultimately influencing the opinions of the other players. Therefore, the key takeaway is that opinions are a viable and effective channel for achieving collective change, and raising awareness using committed minority holds substantial promise as a method of choice.



\section{Perspectives}


The control methods discussed in this article have been primarily open-loop or static, which has advantages in terms of being easier to analyze, but also several disadvantages. For instance, committed minority are assumed to permanently act as such and efforts to increase sensitivity to trends or visibility are assumed to remain always active. This may consume resources unnecessarily. 
Moreover, uncertainty and noise are major issues for social systems due to limited knowledge of the exact game structure, and inherent complexity and stochasticity of human decision-making. 
An emerging direction is the development of closed-loop schemes that employ the power of feedback control to overcome these limitations. Of particular interest are adaptive-gain schemes, which dynamically adjust the payoff terms in a game and requires only limited information on the payoff structure~\cite{MartinezPiazuelo2022b,Zino2025tac}. Here,  key open questions include the design of dynamic targeting control that adjusts which nodes the intervention is applied to, and the impact of revision protocols that reflect humans' bounded rationality and stochasticity in decision making.

In this article, we have not explicitly considered the information environment under which the social dynamics are occurring, but there may be specific avenues that emerge if we consider socio-technical systems. In fact, much of our daily information sharing or learning occurs in online environments, such as social media platforms. Recent studies have examined the role of recommender systems, algorithms that determine what information is visible on a platform, on shaping user opinions~\cite{Jannach2010,Rossi2022,Sprenger2024}. An important direction is to examine whether interventions can be made more efficient and effective if delivered using recommender systems, helping to raise awareness among those who would most benefit from it. More broadly speaking, active consideration of the technological features of the information environment to increase awareness is part of the growing literature on cyber-physical-human systems~\cite{bookcphs2023,Cao2025}.

Furthermore, it is important to remember that interventions on real-world socio-technical systems require careful considerations. Ensuring that interventions and algorithms maintain fairness for everyone is a central theme of this special issue, and it may need to be considered when designing interventions that raise awareness. In particular, it may be necessary to consider a tradeoff between interventions that reduces the resources needed, against those that promote awareness to the most marginalized or impacted segments of society~\cite{Villa2025}. Another important consideration is that policymakers must often achieve their objective given limited budget and resources, paving the way for optimization techniques to be employed. A particularly interesting question emerges when we observe that raising awareness of behavior or of opinions may involve different resources and have different costs.

The future work discussed so far sits squarely within the scope of the systems and control community. However, modeling collective human behavior raises interdisciplinary challenges that requires bridging the gap to other scientific communities. A major challenge is integrating empirical data into the models to calibrate them, and there are at least two central issues here. First, some functional forms can be difficult to be identified (for example, the return function of a public goods game if it is nonlinear). In these scenarios, approaches borrowed from physics-informed machine learning~\cite{Karniadakis2021} could pave the way for the development of socially-informed data-driven control techniques. Second, decision-making and social processes can be highly context-dependent. While the underlying mechanisms might be generalizable, specific processes that are relevant to protest movements may not appear in the adoption of sustainable behaviors or norms. In other words, the models developed and data collected in one context may not be applicable (or at least applicable to the same degree) for another context, which underpins the need to collaborate with other disciplines for model development and data integration.

Another challenge is with the data collection itself. Data can be obtained from small-scale experiments involving real human participants, or through large-scale datasets obtained via surveys or social media. Both have different advantages and disadvantages, and developing closer collaborations with scientists in these domains would increase the quality and applicability of the data collected. An newly established approach to tackle data scarcity is the use of generate synthetic data as a replacement for, or supplement to, human generated data. Preliminary studies indicate that such synthetic data can closely approximate those from real experiments, especially using the emerging Large Language Models (LLMs)~\cite{VanderStigchel2026,Marchi2026}.

Beyond data integration, we highlight the possibility of ``closing-the-loop,'' using the aforementioned models to inform real-world experiments and field studies, which are often expensive and logistically challenging to conduct. Accurately calibrated models can be used to test counterfactuals and explore different intervention scenarios to identify the most promising  options, which can then be tested in the real world. This approach has been explored in urban planning and promotion of sustainable behavior~\cite{Hoffmann2024,Farjadnia2026}, but there remains other unexplored applications. Experiments and field studies are infeasible in some contexts, such as protest movements that involve radical and violent tactics or repressive authorities, and thus models may be  pivotal in hypothesis testing and extending existing theories. Looking to the future, there are many opportunities to meaningfully engage with scientific communities that are not fully aware of the unique advantages of systems- and control-theoretic methods, including dynamics, feedback, and optimization, helping to solve societal problems for the collective good.

\section{Acknowledgments}

The authors thank Ella Connor Davidson and Roberta Raineri for their help with simulations. M. Ye was supported by the Australian Government through the Australian Research Council (DE250100199) and the Office of National Intelligence (NI240100203).

\section{Author Information}

\begin{IEEEbiography}{{M}engbin Ye}{\,}(ben.ye@adelaide.edu.au) received the B.E. degree (First Class Honours) in Mechanical Engineering from the University of Auckland, New Zealand (2013), and the Ph.D.  in Engineering at the Australian National University (2018). He is a Senior Lecturer with the Adelaide Data Science Centre, University of Adelaide, Australia, since 2025, holding an Australian Research Council Discovery Early Career Research Award (DECRA). He was awarded the J.G. Crawford Prize (Interdisciplinary) in 2018, ANU’s premier award recognizing graduate research excellence, and the 2018 Springer PhD Thesis Prize. His research interests include opinion formation and decision making in complex social networks, epidemic modeling and control, and cooperative control of multi-agent
systems. He is a Member of IEEE.
\end{IEEEbiography}

\begin{IEEEbiography}{Lorenzo Zino}{\,}(lorenzo.zino@polito.it) received the B.S. (2012), M.S. (summa cum laude, 2014), and Ph.D. (with honors, 2018) in Applied Mathematics  from  Politecnico  di  Torino, Italy. He is an Assistant Professor with the Department of Electronics and Telecommunications, Politecnico di Torino, Italy, since 2022. In 2024, he received the Best Young Author Journal Paper Award from the  IEEE CSS Italy Chapter. He is Associate Editor of {Scientific Reports}, {International Journal of Control}, and {IEEE Control Systems Letters}, and the  IEEE CSS and EUCA Conference Editorial Boards. His research interests include modeling, analysis, and control of dynamics over networks, social dynamics, and game theory. He is a Senior Member of IEEE. 
\end{IEEEbiography}

\begin{IEEEbiography}{Ming Cao}{\,}(m.cao@rug.nl) received  the B.E. (1999) and M.E. (2002) from Tsinghua University, China, and the Ph.D. degree in 2007 from Yale University, USA. He is a Professor of networks and robotics with the Engineering and Technology Institute (ENTEG) at the University of Groningen, the Netherlands, since 2016. Since 2022 he is the director of the Jantina Tammes School of Digital Society, Technology and AI at the same university. He received the Manfred Thoma medal from IFAC (2017) and the European Control Award  from EUCA (2016).  He is a Senior Editor for {Systems \& Control Letters}. His research interests include autonomous robots and multi-agent systems, complex networks and decision-making processes. He is a fellow of IEEE, a Member of the IFAC Council and a vice chair of the IFAC Technical Committee on Large-Scale Complex Systems. 
\end{IEEEbiography}

\bibliographystyle{IEEEtran}
\bibliography{MYE_ANU}

\endarticle

\end{document}

%% file: fig/coord.tex
\begin {tikzpicture}
\usetikzlibrary{shapes.geometric}
\tikzstyle{peers}=[draw,circle, text=black,  fill=white,inner sep=0pt, minimum size=.6cm]
\draw [draw=none,ultra thick,fill=gray!30,rounded corners=3] (-2.8,2.8) rectangle (5.7,6.8); 
\node[peers,fill=cyan,cyan] (1) at (-0.9,4) {\color{gray!30}+};
\node[peers,fill=red,red] (2) at (1.6,3.6) {\color{gray!30}-};
\node[peers,fill=red,red] (3) at (4.2,4) {\color{gray!30}-};

\node[peers,fill=cyan,cyan] (4) at (-.2,6) {\color{gray!30}+};
\node[peers,minimum size=.9cm,fill=gray!20,very thick,draw=gray] (5) at (2.2,5.4) {\color{gray}\large $i$};
\node[peers,fill=red,red] (6) at (4,6) {\color{gray!30}-};
\node (t) at (4,3) {\,};

\foreach \i/\j in {1/5,4/5}
{\path[gray] (\i) edge node[fill=gray!30] {\color{cyan}$1+\alpha$} (\j) ;}

\foreach \i/\j in {3/5,6/5,2/5}
{\path[gray] (\i) edge node[fill=gray!30]{\color{red}$1$} (\j) ;}

\foreach \i/\j in {1/4,2/3,1/2}
{\path[gray] (\i) edge node{} (\j) ;}

\path[gray, dashed] (1) edge node {} (-1.9,5) ;
\path[gray, dashed] (1) edge node {} (-2.4,3.8) ;
\path[gray, dashed] (4) edge node {} (-2.1,5.5) ;
\path[gray, dashed] (3) edge node {} (5.3,3.9) ;
\path[gray, dashed] (6) edge node {} (5.5,5.6) ;

\end{tikzpicture}

%% file: fig/coord_ne.tex
\begin {tikzpicture}
\usetikzlibrary{shapes.geometric}
\tikzstyle{peers}=[draw,circle, text=black,  fill=white,inner sep=0pt, minimum size=.5cm]
\node[peers,fill=red,red] (1) at (0,0) {\color{white}-};
\node[peers,fill=red,red] (2) at (1,0.6) {\color{white}-};
\node[peers,fill=red,red] (3) at (0,1.2) {\color{white}-};

\node[peers,fill=red,red] (4) at (3.2,0) {\color{white}-};
\node[peers,fill=red,red] (5) at (2.2,0.6) {\color{white}-};
\node[peers,fill=red,red] (6) at (3.2,1.2) {\color{white}-};

\foreach \i/\j in {1/2,1/3,2/3,4/5,4/6,5/6,2/5}
{\path[gray] (\i) edge node {} (\j) ;}

\node[peers,fill=cyan,cyan] (11) at (0,-2) {\color{white}+};
\node[peers,fill=cyan,cyan] (21) at (1,-1.4) {\color{white}+};
\node[peers,fill=cyan,cyan] (31) at (0,-0.8) {\color{white}+};

\node[peers,fill=cyan,cyan] (41) at (3.2,-2) {\color{white}+};
\node[peers,fill=cyan,cyan] (51) at (2.2,-1.4) {\color{white}+};
\node[peers,fill=cyan,cyan] (61) at (3.2,-0.8) {\color{white}+};

\foreach \i/\j in {11/21,11/31,21/31,41/51,41/61,51/61,21/51}
{\path[gray] (\i) edge node {} (\j) ;}
\end{tikzpicture}\quad
\begin {tikzpicture}
\usetikzlibrary{shapes.geometric}
\tikzstyle{peers}=[draw,circle, text=black,  fill=white,inner sep=0pt, minimum size=.5cm]
\node[peers,fill=cyan,cyan] (1) at (0,0) {\color{white}+};
\node[peers,fill=cyan,cyan] (2) at (1,0.6) {\color{white}+};
\node[peers,fill=cyan,cyan] (3) at (0,1.2) {\color{white}+};

\node[peers,fill=red,red] (4) at (3.2,0) {\color{white}-};
\node[peers,fill=red,red] (5) at (2.2,0.6) {\color{white}-};
\node[peers,fill=red,red] (6) at (3.2,1.2) {\color{white}-};

\foreach \i/\j in {1/2,1/3,2/3,4/5,4/6,5/6,2/5}
{\path[gray] (\i) edge node {} (\j) ;}

\node[peers,fill=red,red] (11) at (0,-2) {\color{white}-};
\node[peers,fill=red,red] (21) at (1,-1.4) {\color{white}-};
\node[peers,fill=red,red] (31) at (0,-0.8) {\color{white}-};

\node[peers,fill=cyan,cyan] (41) at (3.2,-2) {\color{white}+};
\node[peers,fill=cyan,cyan] (51) at (2.2,-1.4) {\color{white}+};
\node[peers,fill=cyan,cyan] (61) at (3.2,-0.8) {\color{white}+};

\foreach \i/\j in {11/21,11/31,21/31,41/51,41/61,51/61,21/51}
{\path[gray] (\i) edge node {} (\j) ;}
\end{tikzpicture}

%% file: fig/pgg2.tex
\begin{tikzpicture}
    \tikzstyle{Qmatrix}=[rectangle,rounded corners=3,fill=orange,minimum size=22pt,inner sep=5pt];

\draw [draw=none,ultra thick,fill=gray!30,rounded corners=3] (-3.5,-2) rectangle (3.5,2); 

\foreach \x/\y/\n in {0/1.5/1,2.4/0.7/2,0/-1.5/4}
{\node[draw=red,  inner sep=0pt, circle, minimum size=18pt,fill=red] (\n) at  (\x,\y) {\small{\textcolor{white}{$-$}}};}

\foreach \x/\y/\n in {2.4/-.7/5,-2.4/-0.7/3,-2.4/0.7/6}
{\node[draw=cyan,  inner sep=0pt, circle, minimum size=18pt,fill=cyan] (\n) at  (\x,\y) {\small{\textcolor{white}{$+$}}};}

  \node[Qmatrix] (A) at (0,0) {\color{gray!30}pool};

  \path[->,thick,dotted, bend right=15] (3) edge node[above] {}  (A) ;    
  \path[->,thick,dotted, bend right=15] (6) edge node[above] {}  (A) ;    
  \path[->,thick,dotted, bend right=15] (5) edge node[above] {}  (A) ;

\foreach \n in {1,4,2}
{\path[->,thick] (A) edge node[above] {}  (\n) ;} 
\foreach \n in {3,6,5}
{\path[->,thick,bend right=15] (A) edge node[right] {}  (\n) ;}  
\end{tikzpicture}

%% file: fig/trend1.tex
\definecolor{myviolet}{RGB}{72,160,66}
\definecolor{myblue}{RGB}{20,155,204}
\definecolor{myyellow}{RGB}{190,170,0}
\definecolor{myblue}{RGB}{0,66,255}\scalebox{0.85}{\begin{tikzpicture}
\draw [draw=none,ultra thick,fill=orange!30,rounded corners=3] (-.8,-.6) rectangle (0.3,1.3); 

        \draw [draw=none,ultra thick,fill=myviolet!30,rounded corners=3] (-2.1,1.5) rectangle (2,2.5); 
      \draw [draw=none,ultra thick,fill=myviolet!30,rounded corners=3] (0.5,-.6) rectangle (2.1,2.5); 
          \draw [draw=none,ultra thick,fill=myviolet!30,rounded corners=3] (-2.1,-.6) rectangle (-1,2); 

\foreach \x/\y/\n in {-.1/0/1,1.7/.3/4,-.4/1/5,-1.5/.75/6,-1.2/2/7,.2/2.2/8}
{\node[draw=red,  inner sep=0pt, circle, minimum size=10pt,fill=red] (\n) at  (\x,\y) {\small{\textcolor{white}{$-$}}};}

\foreach \x/\y/\n in {.9/-.25/2,1.1/1.5/3}
{\node[draw=cyan,  inner sep=0pt, circle, minimum size=10pt,fill=cyan] (\n) at  (\x,\y) {\small{\textcolor{white}{$+$}}};}

 \tikzset{mystyle/.style={->,black}} 
 \tikzset{every node/.style={fill=white}}

\foreach  \i/\j in {2/1,2/5,2/3,3/1,3/7,4/2,7/5,8/3,8/2}
{\path (\i) edge[mystyle]  (\j) ;}

\foreach  \i/\j in {3/4,4/3,6/7,7/6,7/8,8/7,6/5}
{\path (\i) edge[mystyle, bend right=15]  (\j) ;}

\foreach  \i/\j in {6/5}
{\path (\i) edge[mystyle, bend left=15]  (\j) ;}

\foreach  \i/\j in {4/8}
{\path (\i) edge[mystyle, bend right=45]  (\j) ;}


\end{tikzpicture}}

%% file: fig/trend2.tex
\definecolor{myviolet}{RGB}{72,160,66}
\definecolor{myblue}{RGB}{20,155,204}
\definecolor{myyellow}{RGB}{190,170,0}
\definecolor{myblue}{RGB}{0,66,255}\scalebox{0.85}{\begin{tikzpicture}
\draw [draw=none,ultra thick,fill=orange!30,rounded corners=3] (-.8,-.6) rectangle (0.3,1.3); 

        \draw [draw=none,ultra thick,fill=myviolet!30,rounded corners=3] (-2.1,1.5) rectangle (2,2.5); 
      \draw [draw=none,ultra thick,fill=myviolet!30,rounded corners=3] (0.5,-.6) rectangle (2.1,2.5); 
          \draw [draw=none,ultra thick,fill=myviolet!30,rounded corners=3] (-2.1,-.6) rectangle (-1,2); 

\foreach \x/\y/\n in {.9/-.25/2,1.1/1.5/3,-1.5/.75/6,-1.2/2/7}
{\node[draw=red,  inner sep=0pt, circle, minimum size=10pt,fill=red] (\n) at  (\x,\y) {\small{\textcolor{white}{$-$}}};}

\foreach \x/\y/\n in {1.7/.3/4,-.1/0/1,-.4/1/5,.2/2.2/8}
{\node[draw=cyan,  inner sep=0pt, circle, minimum size=10pt,fill=cyan] (\n) at  (\x,\y) {\small{\textcolor{white}{$+$}}};}

\tikzset{mystyle/.style={->,black}} 
\tikzset{every node/.style={fill=white}}

\foreach  \i/\j in {2/1,2/5,2/3,3/1,3/7,4/2,7/5,8/3,8/2}
{\path (\i) edge[mystyle]  (\j) ;}

\foreach  \i/\j in {3/4,4/3,6/7,7/6,7/8,8/7,6/5}
{\path (\i) edge[mystyle, bend right=15]  (\j) ;}

\foreach  \i/\j in {6/5}
{\path (\i) edge[mystyle, bend left=15]  (\j) ;}

\foreach  \i/\j in {4/8}
{\path (\i) edge[mystyle, bend right=45]  (\j) ;}

\end{tikzpicture}}

%% file: fig/nexus1.tex
\begin{tikzpicture}
\begin{axis}[%
 axis lines=left,
 x   axis line style={-},
  y   axis line style={-},
width=2.5cm,
height=2.8cm,
scale only axis,
xmin=-1,
xmax=1.98,
xlabel={advantage, $\alpha$},
ylabel={sensitivity, $u_t$},
ymin=0,
ymax=1,
axis background/.style={fill=white},
]

\addplot[ultra thick, black,name path = A]
  table[row sep=crcr]{%
-1	0.6667\\
-0.5	0.6667\\
-0.5	0.1111\\
};

\addplot[ultra thick, black,name path = D]
  table[row sep=crcr]{%
-0.5	0.1111\\
1	0.1111\\
1	0\\
6	0\\
};

\addplot[draw=none, black,name path = B]
  table[row sep=crcr]{%
-1	0\\
-0.5	0\\
};

\addplot[draw=none, black,name path = C]
  table[row sep=crcr]{%
-0.5	0\\
6	0\\
};
\addplot[draw=none, black,name path = H]
  table[row sep=crcr]{%
-1	1\\
6	1\\
};

    \addplot[cyan!50] fill between[of=H and A];
        \addplot[cyan!50] fill between[of=H and D];
          \addplot[gray!50] fill between[of=A and B];

      \addplot[gray!50] fill between[of=D and C];

\end{axis}
\end{tikzpicture}

%% file: fig/nexus2.tex
\begin{tikzpicture}
\begin{axis}[%
 axis lines=left,
 x   axis line style={-},
  y   axis line style={-},
width=2.5 cm,
height=2.8 cm,
scale only axis,
xmin=0,
xmax=.129,
 extra y ticks ={0},
    extra y tick labels={$0$},
     extra x ticks ={0},
     extra x tick labels={$0$},
xlabel={sensitivity, $u_t$},
ylabel={innovators, $\zeta_0$},
xticklabel style={/pgf/number format/fixed},  
ymin=0.19,
ymax=.5,
axis background/.style={fill=white},
]

\addplot[ultra thick, cyan!50,name path = A]
  table[row sep=crcr]{%
0	0.5\\
0.000371609067261241	0.499627975984368\\
0.000743218134522482	0.499255119387188\\
0.00111482720178372	0.498881426156542\\
0.00148643626904496	0.498506892207252\\
0.00185804533630621	0.498131513420495\\
0.00222965440356745	0.497755285643406\\
0.00260126347082869	0.497378204688676\\
0.00297287253808993	0.497000266334146\\
0.00334448160535117	0.496621466322392\\
0.00371609067261241	0.496241800360306\\
0.00408769973987365	0.495861264118669\\
0.00445930880713489	0.495479853231717\\
0.00483091787439614	0.495097563296701\\
0.00520252694165738	0.494714389873444\\
0.00557413600891862	0.494330328483879\\
0.00594574507617986	0.493945374611597\\
0.0063173541434411	0.493559523701373\\
0.00668896321070234	0.49317277115869\\
0.00706057227796358	0.492785112349261\\
0.00743218134522482	0.492396542598531\\
0.00780379041248606	0.492007057191184\\
0.0081753994797473	0.491616651370633\\
0.00854700854700855	0.491225320338504\\
0.00891861761426979	0.490833059254115\\
0.00929022668153103	0.490439863233945\\
0.00966183574879227	0.490045727351087\\
0.0100334448160535	0.489650646634705\\
0.0104050538833148	0.489254616069475\\
0.010776662950576	0.488857630595013\\
0.0111482720178372	0.4884596851053\\
0.0115198810850985	0.488060774448096\\
0.0118914901523597	0.487660893424343\\
0.012263099219621	0.487260036787558\\
0.0126347082868822	0.486858199243216\\
0.0130063173541434	0.486455375448124\\
0.0133779264214047	0.48605156000978\\
0.0137495354886659	0.485646747485729\\
0.0141211445559272	0.485240932382899\\
0.0144927536231884	0.484834109156931\\
0.0148643626904496	0.484426272211499\\
0.0152359717577109	0.484017415897613\\
0.0156075808249721	0.48360753451291\\
0.0159791898922334	0.483196622300942\\
0.0163507989594946	0.482784673450437\\
0.0167224080267559	0.482371682094561\\
0.0170940170940171	0.481957642310157\\
0.0174656261612783	0.481542548116976\\
0.0178372352285396	0.481126393476891\\
0.0182088442958008	0.480709172293098\\
0.0185804533630621	0.480290878409308\\
0.0189520624303233	0.479871505608914\\
0.0193236714975845	0.479451047614151\\
0.0196952805648458	0.479029498085239\\
0.020066889632107	0.478606850619509\\
0.0204384986993683	0.478183098750514\\
0.0208101077666295	0.477758235947124\\
0.0211817168338907	0.477332255612602\\
0.021553325901152	0.476905151083668\\
0.0219249349684132	0.47647691562954\\
0.0222965440356745	0.47604754245096\\
0.0226681531029357	0.475617024679201\\
0.023039762170197	0.475185355375055\\
0.0234113712374582	0.474752527527806\\
0.0237829803047194	0.474318534054173\\
0.0241545893719807	0.473883367797248\\
0.0245261984392419	0.473447021525397\\
0.0248978075065032	0.473009487931157\\
0.0252694165737644	0.472570759630096\\
0.0256410256410256	0.47213082915966\\
0.0260126347082869	0.471689688977996\\
0.0263842437755481	0.471247331462753\\
0.0267558528428094	0.470803748909854\\
0.0271274619100706	0.470358933532252\\
0.0274990709773318	0.469912877458651\\
0.0278706800445931	0.469465572732214\\
0.0282422891118543	0.469017011309236\\
0.0286138981791156	0.468567185057789\\
0.0289855072463768	0.46811608575635\\
0.029357116313638	0.467663705092393\\
0.0297287253808993	0.46721003466095\\
0.0301003344481605	0.46675506596315\\
0.0304719435154218	0.466298790404727\\
0.030843552582683	0.465841199294489\\
0.0312151616499443	0.465382283842766\\
0.0315867707172055	0.464922035159816\\
0.0319583797844667	0.464460444254208\\
0.032329988851728	0.463997502031159\\
0.0327015979189892	0.463533199290849\\
0.0330732069862505	0.463067526726685\\
0.0334448160535117	0.462600474923544\\
0.0338164251207729	0.462132034355964\\
0.0341880341880342	0.461662195386309\\
0.0345596432552954	0.461190948262883\\
0.0349312523225567	0.460718283118006\\
0.0353028613898179	0.460244189966058\\
0.0356744704570792	0.459768658701462\\
0.0360460795243404	0.459291679096638\\
0.0364176885916016	0.458813240799905\\
0.0367892976588629	0.458333333333333\\
0.0371609067261241	0.457851946090557\\
0.0375325157933854	0.457369068334529\\
0.0379041248606466	0.45688468919523\\
0.0382757339279078	0.456398797667322\\
0.0386473429951691	0.455911382607755\\
0.0390189520624303	0.455422432733306\\
0.0393905611296916	0.454931936618075\\
0.0397621701969528	0.454439882690915\\
0.040133779264214	0.4539462592328\\
0.0405053883314753	0.453451054374139\\
0.0408769973987365	0.452954256092022\\
0.0412486064659978	0.452455852207397\\
0.041620215533259	0.451955830382187\\
0.0419918246005202	0.451454178116334\\
0.0423634336677815	0.45095088274477\\
0.0427350427350427	0.450445931434318\\
0.043106651802304	0.449939311180517\\
0.0434782608695652	0.449431008804366\\
0.0438498699368265	0.448921010948991\\
0.0442214790040877	0.448409304076226\\
0.0445930880713489	0.447895874463117\\
0.0449646971386102	0.447380708198323\\
0.0453363062058714	0.446863791178446\\
0.0457079152731327	0.446345109104248\\
0.0460795243403939	0.445824647476787\\
0.0464511334076551	0.445302391593447\\
0.0468227424749164	0.444778326543864\\
0.0471943515421776	0.444252437205751\\
0.0475659606094389	0.443724708240613\\
0.0479375696767001	0.443195124089344\\
0.0483091787439613	0.442663668967716\\
0.0486807878112226	0.442130326861742\\
0.0490523968784838	0.441595081522921\\
0.0494240059457451	0.441057916463347\\
0.0497956150130063	0.44051881495069\\
0.0501672240802676	0.439977760003045\\
0.0505388331475288	0.439434734383634\\
0.05091044221479	0.438889720595361\\
0.0512820512820513	0.438342700875221\\
0.0516536603493125	0.43779365718855\\
0.0520252694165738	0.437242571223116\\
0.052396878483835	0.436689424383039\\
0.0527684875510962	0.436134197782542\\
0.0531400966183575	0.435576872239523\\
0.0535117056856187	0.435017428268936\\
0.05388331475288	0.434455846075993\\
0.0542549238201412	0.433892105549152\\
0.0546265328874024	0.433326186252913\\
0.0549981419546637	0.432758067420389\\
0.0553697510219249	0.432187727945668\\
0.0557413600891862	0.431615146375935\\
0.0561129691564474	0.431040300903365\\
0.0564845782237087	0.430463169356767\\
0.0568561872909699	0.429883729192967\\
0.0572277963582311	0.429301957487939\\
0.0575994054254924	0.428717830927646\\
0.0579710144927536	0.428131325798606\\
0.0583426235600149	0.427542417978159\\
0.0587142326272761	0.426951082924419\\
0.0590858416945373	0.426357295665921\\
0.0594574507617986	0.425761030790918\\
0.0598290598290598	0.425162262436343\\
0.0602006688963211	0.424560964276411\\
0.0605722779635823	0.423957109510842\\
0.0609438870308436	0.4233506708527\\
0.0613154960981048	0.422741620515822\\
0.061687105165366	0.422129930201831\\
0.0620587142326273	0.421515571086704\\
0.0624303232998885	0.420898513806888\\
0.0628019323671497	0.420278728444937\\
0.063173541434411	0.419656184514652\\
0.0635451505016722	0.419030850945703\\
0.0639167595689335	0.418402696067712\\
0.0642883686361947	0.417771687593771\\
0.064659977703456	0.417137792603376\\
0.0650315867707172	0.416500977524738\\
0.0654031958379784	0.415861208116464\\
0.0657748049052397	0.415218449448559\\
0.0661464139725009	0.414572665882734\\
0.0665180230397622	0.413923821051976\\
0.0668896321070234	0.413271877839357\\
0.0672612411742847	0.412616798356042\\
0.0676328502415459	0.411958543918456\\
0.0680044593088071	0.411297075024575\\
0.0683760683760684	0.410632351329295\\
0.0687476774433296	0.409964331618844\\
0.0691192865105908	0.409292973784175\\
0.0694908955778521	0.40861823479331\\
0.0698625046451133	0.407940070662562\\
0.0702341137123746	0.407258436426597\\
0.0706057227796358	0.406573286107276\\
0.0709773318468971	0.405884572681199\\
0.0713489409141583	0.405192248045904\\
0.0717205499814195	0.404496262984646\\
0.0720921590486808	0.403796567129676\\
0.072463768115942	0.403093108923949\\
0.0728353771832033	0.402385835581171\\
0.0732069862504645	0.401674693044107\\
0.0735785953177258	0.400959625941044\\
0.073950204384987	0.400240577540313\\
0.0743218134522482	0.399517489702768\\
0.0746934225195095	0.398790302832098\\
0.0750650315867707	0.398058955822865\\
0.075436640654032	0.397323386006121\\
0.0758082497212932	0.396583529092481\\
0.0761798587885544	0.395839319112493\\
0.0765514678558157	0.395090688354155\\
0.0769230769230769	0.394337567297406\\
0.0772946859903382	0.39357988454541\\
0.0776662950575994	0.392817566752435\\
0.0780379041248606	0.39205053854813\\
0.0784095131921219	0.391278722457965\\
0.0787811222593831	0.390502038819601\\
0.0791527313266444	0.389720405694925\\
0.0795243403939056	0.388933738777485\\
0.0798959494611668	0.38814195129501\\
0.0802675585284281	0.38734495390671\\
0.0806391675956893	0.386542654594995\\
0.0810107766629506	0.385734958551242\\
0.0813823857302118	0.384921768055216\\
0.0817539947974731	0.384102982347689\\
0.0821256038647343	0.383278497495796\\
0.0824972129319955	0.382448206250608\\
0.0828688219992568	0.381611997896375\\
0.083240431066518	0.380769758090817\\
0.0836120401337793	0.379921368695822\\
0.0839836492010405	0.379066707597823\\
0.0843552582683017	0.378205648517093\\
0.084726867335563	0.377338060805079\\
0.0850984764028242	0.376463809228882\\
0.0854700854700855	0.375582753741838\\
0.0858416945373467	0.374694749239111\\
0.0862133036046079	0.373799645297073\\
0.0865849126718692	0.372897285895135\\
0.0869565217391304	0.371987509118567\\
0.0873281308063917	0.371070146840689\\
0.0876997398736529	0.370145024382662\\
0.0880713489409142	0.369211960148899\\
0.0884429580081754	0.368270765235956\\
0.0888145670754366	0.367321243012466\\
0.0891861761426979	0.366363188667493\\
0.0895577852099591	0.365396388724311\\
0.0899293942772204	0.36442062051635\\
0.0903010033444816	0.363435651621629\\
0.0906726124117428	0.362441239251575\\
0.0910442214790041	0.361437129589673\\
0.0914158305462653	0.36042305707479\\
0.0917874396135266	0.359398743623421\\
0.0921590486807878	0.358363897784368\\
0.092530657748049	0.357318213818527\\
0.0929022668153103	0.356261370695513\\
0.0932738758825715	0.355193030997742\\
0.0936454849498328	0.354112839721304\\
0.094017094017094	0.353020422961517\\
0.0943887030843552	0.351915386469263\\
0.0947603121516165	0.350797314062268\\
0.0951319212188777	0.34966576587306\\
0.095503530286139	0.348520276412628\\
0.0958751393534002	0.347360352425483\\
0.0962467484206615	0.346185470507983\\
0.0966183574879227	0.344995074457169\\
0.096989966555184	0.343788572311856\\
0.0973615756224452	0.342565333041161\\
0.0977331846897064	0.341324682827672\\
0.0981047937569677	0.340065900882932\\
0.0984764028242289	0.338788214721166\\
0.0988480118914901	0.337490794802983\\
0.0992196209587514	0.336172748443214\\
0.0995912300260126	0.334833112855366\\
0.0999628390932739	0.333470847178245\\
0.100334448160535	0.332084823296474\\
0.100706057227796	0.330673815224086\\
0.101077666295058	0.329236486766235\\
0.101449275362319	0.327771377104782\\
0.10182088442958	0.326276883863975\\
0.102192493496841	0.324751243095823\\
0.102564102564103	0.32319250547114\\
0.102935711631364	0.321598507758005\\
0.103307320698625	0.319966838394539\\
0.103678929765886	0.318294795588519\\
0.104050538833148	0.316579335859564\\
0.104422147900409	0.314817010215514\\
0.10479375696767	0.313003884123877\\
0.105165366034931	0.311135435945442\\
0.105536975102192	0.309206426289727\\
0.105908584169454	0.307210727417774\\
0.106280193236715	0.305141096657036\\
0.106651802303976	0.30298886958127\\
0.107023411371237	0.300743535224117\\
0.107395020438499	0.29839213263795\\
0.10776662950576	0.295918367346939\\
0.108138238573021	0.293301270189222\\
0.108509847640282	0.290513070467436\\
0.108881456707544	0.287515634772412\\
0.109253065774805	0.284254075335591\\
0.109624674842066	0.280644169241704\\
0.109996283909327	0.276544168886573\\
0.110367892976589	0.271677749238103\\
0.11073950204385	0.265331685928084\\
0.111111111111111	0.25\\
0.111111111111111	0\\};

\addplot[ultra thick, cyan!50,name path = B]
  table[row sep=crcr]{%
0 0\\
0.111111111111111 0\\
};

\addplot[ultra thick, cyan!50,name path = C]
  table[row sep=crcr]{%
0 0.5\\
0.111111111111111 0.5\\
};

\addplot[ultra thick, cyan!50,name path = D]
  table[row sep=crcr]{%
0.1 0.5\\
0.129 0.5\\
};

\addplot[ultra thick, cyan!50,name path = E]
  table[row sep=crcr]{%
0.1 0\\
0.129 0\\
};

   \addplot[cyan!50] fill between[of=A and C];
\addplot[cyan!50] fill between[of=D and E];
     \addplot[red!50] fill between[of=A and B]; 
\end{axis}
\end{tikzpicture}

%% file: fig/consensus.tex
\hspace{-.2cm}\begin{tikzpicture}
\definecolor{mycolor1}{RGB}{72,160,66}
\definecolor{mycolor2}{RGB}{20,155,204}
\definecolor{mycolor3}{RGB}{190,170,0}
\definecolor{mycolor4}{RGB}{195,8,255}
\definecolor{mycolor5}{RGB}{204,0,0}
\definecolor{mycolor6}{RGB}{0,66,255}
\begin{axis}[%
axis lines=left,
x    axis line style={->},
y    axis line style={->},
width=6 cm,
height=3 cm,
scale only axis,
xmin=0,
xmax=38,
ymin=-4,
ymax=3,
ytick={-4,-2,0,2},
yticklabels={$-1$, $-0.5$, $0$, $0.5$},
ylabel={opinion, $y_i(t)$},
xlabel={time, $t$},
axis background/.style={fill=white},
]
\addplot [very thick, color=mycolor1]
  table[row sep=crcr]{%
0	-1.58571932782615\\
0.289191906301376	-1.55655086598371\\
1.73515143780825	-1.41792561022762\\
4.4097184891502	-1.21306272215682\\
7.30073646343245	-1.07110498091422\\
10.8423435907377	-0.979236060782603\\
15.0824405042809	-0.935450037422354\\
19.9290561630079	-0.923092890545719\\
25.6843599211593	-0.924182059890962\\
32.6582208823199	-0.928579682753798\\
40.6582208823199	-0.931725734343875\\
48.6582208823199	-0.933093035050594\\
56.6582208823199	-0.933624057675792\\
64.6582208823199	-0.933819714391414\\
72.6582208823199	-0.933889824104868\\
80	-0.933913369060764\\
};

\addplot [very thick, color=mycolor2]
  table[row sep=crcr]{%
0	-1.27429887375026\\
0.289191906301376	-1.28109593056904\\
1.73515143780825	-1.2876901113936\\
4.4097184891502	-1.23752507549549\\
7.30073646343245	-1.1621456001793\\
10.8423435907377	-1.08371020376056\\
15.0824405042809	-1.02057287407185\\
19.9290561630079	-0.979314696176513\\
25.6843599211593	-0.954825620901287\\
32.6582208823199	-0.942094909774838\\
40.6582208823199	-0.936720069804768\\
48.6582208823199	-0.934885669805013\\
56.6582208823199	-0.934257072253798\\
64.6582208823199	-0.934041132012233\\
72.6582208823199	-0.933966837063063\\
80	-0.933942479110213\\
};

\addplot [very thick, color=mycolor3]
  table[row sep=crcr]{%
0	0.494428564484411\\
0.289191906301376	0.398332738462826\\
1.73515143780825	0.00371493450141502\\
4.4097184891502	-0.453271756283833\\
7.30073646343245	-0.710587389112408\\
10.8423435907377	-0.856732439541154\\
15.0824405042809	-0.921943238822045\\
19.9290561630079	-0.941538992744365\\
25.6843599211593	-0.942779271714713\\
32.6582208823199	-0.939059511152951\\
40.6582208823199	-0.936081034379445\\
48.6582208823199	-0.934751273294675\\
56.6582208823199	-0.934228850169553\\
64.6582208823199	-0.93403521905332\\
72.6582208823199	-0.933965602403905\\
80	-0.933942187888649\\
};

\addplot [very thick, color=mycolor4]
  table[row sep=crcr]{%
0	-3.88829562007765\\
0.289191906301376	-3.71480683819742\\
1.73515143780825	-2.99870609614592\\
4.4097184891502	-2.14768455342485\\
7.30073646343245	-1.63414259527232\\
10.8423435907377	-1.30105311179994\\
15.0824405042809	-1.10910123514806\\
19.9290561630079	-1.01185621437112\\
25.6843599211593	-0.964969732449908\\
32.6582208823199	-0.944635422719786\\
40.6582208823199	-0.937260614629934\\
48.6582208823199	-0.935001127944954\\
56.6582208823199	-0.93428187130348\\
64.6582208823199	-0.934046500844595\\
72.6582208823199	-0.933968012298053\\
80	-0.933942754612721\\
};

\addplot [very thick, color=mycolor5]
  table[row sep=crcr]{%
0	-0.842104617283189\\
0.289191906301376	-0.821339782993308\\
1.73515143780825	-0.755835186153832\\
4.4097184891502	-0.733225331337059\\
7.30073646343245	-0.763425696519625\\
10.8423435907377	-0.812571617235552\\
15.0824405042809	-0.859461146113451\\
19.9290561630079	-0.893165196359498\\
25.6843599211593	-0.914497776169351\\
32.6582208823199	-0.926130080489806\\
40.6582208823199	-0.931213461202446\\
48.6582208823199	-0.932986368925989\\
56.6582208823199	-0.933601992794135\\
64.6582208823199	-0.93381519583905\\
72.6582208823199	-0.933888913292573\\
80	-0.9339131531987\\
};

\addplot [very thick, color=mycolor6]
  table[row sep=crcr]{%
0	1.49242308664619\\
0.289191906301376	1.37189389147401\\
1.73515143780825	0.852875281612913\\
4.4097184891502	0.181202650891394\\
7.30073646343245	-0.262160525808779\\
10.8423435907377	-0.57026335468684\\
15.0824405042809	-0.757038256228882\\
19.9290561630079	-0.854598797609425\\
25.6843599211593	-0.902312326680422\\
32.6582208823199	-0.923067180915464\\
40.6582208823199	-0.930565873446175\\
48.6582208823199	-0.932849312785416\\
56.6582208823199	-0.933572943609883\\
64.6582208823199	-0.933809025666028\\
72.6582208823199	-0.933887598644178\\
80	-0.933912843935592\\
};
\end{axis}
\end{tikzpicture}

%% file: fig/schema.tex
\definecolor{mygreen}{RGB}{72,160,66}
\definecolor{myblue}{RGB}{20,155,204}
\definecolor{myyellow}{RGB}{190,170,0}
\definecolor{myviolet}{RGB}{195,8,255}
\definecolor{myred}{RGB}{204,0,0}
\definecolor{myblue}{RGB}{0,66,255}
\begin{tikzpicture}[->,node distance=2cm,every text node part/.style={align=center}]
    \tikzstyle{Qmatrix}=[rectangle,rounded corners=3,fill=green!30,minimum size=22pt,inner sep=5pt];
    \tikzstyle{network node}=[circle,minimum size=8pt,inner sep=0pt, fill=mygreen];
        \tikzstyle{network node2}=[circle,minimum size=8pt,inner sep=0pt, fill=myviolet];

    \tikzstyle{community}=[minimum size=17pt,inner sep=0pt, fill=blue!50];
    
\draw [draw=none,ultra thick,fill=gray!10,rounded corners=3] (0,-2.5) rectangle (2.4,1);            
            \node (tit) at (1.2cm,.75 cm) {\large\textcolor{gray}{network}};
            
\draw [draw=none,ultra thick,fill=cyan!10,rounded corners=3] (2.9,-2.5) rectangle (5.3,1);            
            \node (tit) at (4.1cm,.75 cm) {\large\textcolor{cyan}{dynamics}};
            
\draw [draw=none,ultra thick,fill=red!10,rounded corners=3] (5.8,-2.5) rectangle (8.2,1);            
            \node (tit) at (7cm,.75 cm) {\large\textcolor{red}{state}};
        
        \node[draw=none] (C) at (1.2cm,-0.7) {\scriptsize{influence layer}};
              \node[draw=none] (I) at (1.2cm,-2.2) {\scriptsize{communication layer}};

             \node[Qmatrix,fill=cyan!50] (OD) at (4.1cm,-1) {best  \\ response};

                           \node[Qmatrix,fill=myred!50] (O) at (7cm,-.1) {action $x_i$};
              \node[Qmatrix,fill=red!50] (A) at (7cm,-1.9) {opinion $y_i$};
            
\node[network node,minimum size=10pt,draw=mygreen] at (1.2cm, 0cm) (i) {\color{mygreen}\small{i}};
\node[network node] at (1.7cm, -.35cm) (1) {\color{white}\scriptsize{$x$}};
\node[network node] at (1.9cm, 0.2cm) (2)  {\color{white}\scriptsize{$x$}};
\node[network node] at (0.6cm, 0.3cm) (3)  {\color{white}\scriptsize{$x$}};

\node[network node2,minimum size=10pt,draw=myviolet] at (1.2cm, -1.3cm) (i2) {\color{myviolet}\small{i}};
\node[network node2] at (0.4cm, -1.5cm) (4)  {\color{white}\scriptsize{$y$}};
\node[network node2] at (1.9cm, -1.8cm) (5) {\color{white}\scriptsize{$y$}};
\node[network node2] at (1.3cm, -1.85cm) (7) {\color{white}\scriptsize{$y$}};


\path[-,mygreen!50] (i) edge  (1);
\path[-,mygreen!50] (i) edge  (2);
\path[-,mygreen!50] (i) edge  (3);
\path[-,myviolet!50] (i2) edge  (4);
\path[-,myviolet!50] (i2) edge  (5);
\path[-,myviolet!50] (i2) edge  (7);

\path[thick,-] (OD) edge [above] node {\scriptsize{determines}} (6.5,-1);
\path[thick] (6.5,-1) edge [above] node {} (O);
\path[thick] (6.5,-1) edge [above] node {} (A);
\path[dashed,thick] (2.2,-1.7) edge  [above] node {} (OD);
\path[dashed,thick] (2,-0.2) edge  [right] node {} (OD);
\path[dashed,thick] (6.5,.75) edge[bend right=30] node[above] {} (OD);
\node[draw=none] (446) at (0.6,-.3) {\color{mygreen}$A$};
\node[draw=none] (446) at (0.6,-1.9) {\color{myviolet}$W$};

\end{tikzpicture}\vspace{-.2cm}

%% file: fig/complete.tex
\begin{tikzpicture}
\begin{axis}[%
colormap={mymap}{[1pt] rgb(0pt)=(1,1,1); rgb(127pt)=(0,1,1); rgb(255pt)=(0,.5,.5)},
width=3 cm,
height=3 cm,
scale only axis,
xmin=0.4,
xmax=7,
ymin=0.4,
point meta max=.5,
point meta min=0.2,
ymax=7,
axis background/.style={fill=white},
 axis x line=left,
        axis y line=left,
ylabel={\footnotesize opinion impact, $\beta$},
xlabel={\footnotesize self-consistency, $\lambda$},xmajorgrids,
ymajorgrids,
colorbar,
colorbar style={at={(1.1,1)}, width=.25cm, yticklabel style={
       /pgf/number format/fixed,
       /pgf/number format/precision=5
},},
]

\addplot[%
surf,
shader=interp, draw=none, mesh/rows=34]
table[row sep=crcr, point meta=\thisrow{c}] {%
x	y	c\\
0.4	0.4	0.455\\
0.4	0.6	0.43\\
0.4	0.8	0.41\\
0.4	1	0.395\\
0.4	1.2	0.38\\
0.4	1.4	0.365\\
0.4	1.6	0.355\\
0.4	1.8	0.345\\
0.4	2	0.335\\
0.4	2.2	0.33\\
0.4	2.4	0.32\\
0.4	2.6	0.315\\
0.4	2.8	0.31\\
0.4	3	0.3\\
0.4	3.2	0.295\\
0.4	3.4	0.29\\
0.4	3.6	0.285\\
0.4	3.8	0.285\\
0.4	4	0.28\\
0.4	4.2	0.275\\
0.4	4.4	0.27\\
0.4	4.6	0.27\\
0.4	4.8	0.265\\
0.4	5	0.26\\
0.4	5.2	0.26\\
0.4	5.4	0.255\\
0.4	5.6	0.255\\
0.4	5.8	0.25\\
0.4	6	0.25\\
0.4	6.2	0.245\\
0.4	6.4	0.245\\
0.4	6.6	0.24\\
0.4	6.8	0.24\\
0.4	7	0.235\\
0.4	7.2	0.235\\
0.6	0.4	0.46\\
0.6	0.6	0.44\\
0.6	0.8	0.42\\
0.6	1	0.4\\
0.6	1.2	0.385\\
0.6	1.4	0.37\\
0.6	1.6	0.355\\
0.6	1.8	0.345\\
0.6	2	0.335\\
0.6	2.2	0.325\\
0.6	2.4	0.315\\
0.6	2.6	0.305\\
0.6	2.8	0.3\\
0.6	3	0.29\\
0.6	3.2	0.285\\
0.6	3.4	0.28\\
0.6	3.6	0.275\\
0.6	3.8	0.27\\
0.6	4	0.265\\
0.6	4.2	0.26\\
0.6	4.4	0.255\\
0.6	4.6	0.25\\
0.6	4.8	0.245\\
0.6	5	0.24\\
0.6	5.2	0.235\\
0.6	5.4	0.235\\
0.6	5.6	0.23\\
0.6	5.8	0.225\\
0.6	6	0.225\\
0.6	6.2	0.22\\
0.6	6.4	0.215\\
0.6	6.6	0.215\\
0.6	6.8	0.21\\
0.6	7	0.21\\
0.6	7.2	0.205\\
0.8	0.4	0.465\\
0.8	0.6	0.445\\
0.8	0.8	0.425\\
0.8	1	0.41\\
0.8	1.2	0.39\\
0.8	1.4	0.38\\
0.8	1.6	0.365\\
0.8	1.8	0.35\\
0.8	2	0.34\\
0.8	2.2	0.33\\
0.8	2.4	0.32\\
0.8	2.6	0.31\\
0.8	2.8	0.305\\
0.8	3	0.295\\
0.8	3.2	0.29\\
0.8	3.4	0.28\\
0.8	3.6	0.275\\
0.8	3.8	0.27\\
0.8	4	0.265\\
0.8	4.2	0.26\\
0.8	4.4	0.255\\
0.8	4.6	0.25\\
0.8	4.8	0.245\\
0.8	5	0.24\\
0.8	5.2	0.235\\
0.8	5.4	0.23\\
0.8	5.6	0.225\\
0.8	5.8	0.22\\
0.8	6	0.22\\
0.8	6.2	0.215\\
0.8	6.4	0.21\\
0.8	6.6	0.21\\
0.8	6.8	0.205\\
0.8	7	0.2\\
0.8	7.2	0.2\\
1	0.4	0.47\\
1	0.6	0.45\\
1	0.8	0.435\\
1	1	0.415\\
1	1.2	0.4\\
1	1.4	0.385\\
1	1.6	0.375\\
1	1.8	0.36\\
1	2	0.35\\
1	2.2	0.34\\
1	2.4	0.33\\
1	2.6	0.32\\
1	2.8	0.31\\
1	3	0.305\\
1	3.2	0.295\\
1	3.4	0.29\\
1	3.6	0.28\\
1	3.8	0.275\\
1	4	0.27\\
1	4.2	0.265\\
1	4.4	0.26\\
1	4.6	0.255\\
1	4.8	0.25\\
1	5	0.245\\
1	5.2	0.24\\
1	5.4	0.235\\
1	5.6	0.23\\
1	5.8	0.225\\
1	6	0.22\\
1	6.2	0.22\\
1	6.4	0.215\\
1	6.6	0.21\\
1	6.8	0.21\\
1	7	0.205\\
1	7.2	0.2\\
1.2	0.4	0.475\\
1.2	0.6	0.46\\
1.2	0.8	0.44\\
1.2	1	0.425\\
1.2	1.2	0.41\\
1.2	1.4	0.395\\
1.2	1.6	0.38\\
1.2	1.8	0.37\\
1.2	2	0.36\\
1.2	2.2	0.35\\
1.2	2.4	0.34\\
1.2	2.6	0.33\\
1.2	2.8	0.32\\
1.2	3	0.31\\
1.2	3.2	0.305\\
1.2	3.4	0.295\\
1.2	3.6	0.29\\
1.2	3.8	0.285\\
1.2	4	0.28\\
1.2	4.2	0.27\\
1.2	4.4	0.265\\
1.2	4.6	0.26\\
1.2	4.8	0.255\\
1.2	5	0.25\\
1.2	5.2	0.245\\
1.2	5.4	0.24\\
1.2	5.6	0.235\\
1.2	5.8	0.235\\
1.2	6	0.23\\
1.2	6.2	0.225\\
1.2	6.4	0.22\\
1.2	6.6	0.22\\
1.2	6.8	0.215\\
1.2	7	0.21\\
1.2	7.2	0.205\\
1.4	0.4	0.48\\
1.4	0.6	0.46\\
1.4	0.8	0.445\\
1.4	1	0.43\\
1.4	1.2	0.415\\
1.4	1.4	0.4\\
1.4	1.6	0.39\\
1.4	1.8	0.38\\
1.4	2	0.365\\
1.4	2.2	0.355\\
1.4	2.4	0.345\\
1.4	2.6	0.34\\
1.4	2.8	0.33\\
1.4	3	0.32\\
1.4	3.2	0.315\\
1.4	3.4	0.305\\
1.4	3.6	0.3\\
1.4	3.8	0.29\\
1.4	4	0.285\\
1.4	4.2	0.28\\
1.4	4.4	0.275\\
1.4	4.6	0.27\\
1.4	4.8	0.265\\
1.4	5	0.26\\
1.4	5.2	0.255\\
1.4	5.4	0.25\\
1.4	5.6	0.245\\
1.4	5.8	0.24\\
1.4	6	0.235\\
1.4	6.2	0.23\\
1.4	6.4	0.23\\
1.4	6.6	0.225\\
1.4	6.8	0.22\\
1.4	7	0.22\\
1.4	7.2	0.215\\
1.6	0.4	0.48\\
1.6	0.6	0.465\\
1.6	0.8	0.45\\
1.6	1	0.435\\
1.6	1.2	0.42\\
1.6	1.4	0.41\\
1.6	1.6	0.395\\
1.6	1.8	0.385\\
1.6	2	0.375\\
1.6	2.2	0.365\\
1.6	2.4	0.355\\
1.6	2.6	0.345\\
1.6	2.8	0.34\\
1.6	3	0.33\\
1.6	3.2	0.32\\
1.6	3.4	0.315\\
1.6	3.6	0.31\\
1.6	3.8	0.3\\
1.6	4	0.295\\
1.6	4.2	0.29\\
1.6	4.4	0.285\\
1.6	4.6	0.275\\
1.6	4.8	0.27\\
1.6	5	0.265\\
1.6	5.2	0.26\\
1.6	5.4	0.255\\
1.6	5.6	0.255\\
1.6	5.8	0.25\\
1.6	6	0.245\\
1.6	6.2	0.24\\
1.6	6.4	0.235\\
1.6	6.6	0.23\\
1.6	6.8	0.23\\
1.6	7	0.225\\
1.6	7.2	0.22\\
1.8	0.4	0.485\\
1.8	0.6	0.47\\
1.8	0.8	0.455\\
1.8	1	0.44\\
1.8	1.2	0.43\\
1.8	1.4	0.415\\
1.8	1.6	0.405\\
1.8	1.8	0.39\\
1.8	2	0.38\\
1.8	2.2	0.37\\
1.8	2.4	0.365\\
1.8	2.6	0.355\\
1.8	2.8	0.345\\
1.8	3	0.34\\
1.8	3.2	0.33\\
1.8	3.4	0.325\\
1.8	3.6	0.315\\
1.8	3.8	0.31\\
1.8	4	0.305\\
1.8	4.2	0.295\\
1.8	4.4	0.29\\
1.8	4.6	0.285\\
1.8	4.8	0.28\\
1.8	5	0.275\\
1.8	5.2	0.27\\
1.8	5.4	0.265\\
1.8	5.6	0.26\\
1.8	5.8	0.255\\
1.8	6	0.25\\
1.8	6.2	0.25\\
1.8	6.4	0.245\\
1.8	6.6	0.24\\
1.8	6.8	0.235\\
1.8	7	0.235\\
1.8	7.2	0.23\\
2	0.4	0.485\\
2	0.6	0.47\\
2	0.8	0.46\\
2	1	0.445\\
2	1.2	0.43\\
2	1.4	0.42\\
2	1.6	0.41\\
2	1.8	0.4\\
2	2	0.39\\
2	2.2	0.38\\
2	2.4	0.37\\
2	2.6	0.36\\
2	2.8	0.355\\
2	3	0.345\\
2	3.2	0.335\\
2	3.4	0.33\\
2	3.6	0.325\\
2	3.8	0.315\\
2	4	0.31\\
2	4.2	0.305\\
2	4.4	0.3\\
2	4.6	0.295\\
2	4.8	0.29\\
2	5	0.285\\
2	5.2	0.28\\
2	5.4	0.275\\
2	5.6	0.27\\
2	5.8	0.265\\
2	6	0.26\\
2	6.2	0.255\\
2	6.4	0.25\\
2	6.6	0.25\\
2	6.8	0.245\\
2	7	0.24\\
2	7.2	0.235\\
2.2	0.4	0.485\\
2.2	0.6	0.475\\
2.2	0.8	0.46\\
2.2	1	0.45\\
2.2	1.2	0.435\\
2.2	1.4	0.425\\
2.2	1.6	0.415\\
2.2	1.8	0.405\\
2.2	2	0.395\\
2.2	2.2	0.385\\
2.2	2.4	0.375\\
2.2	2.6	0.37\\
2.2	2.8	0.36\\
2.2	3	0.35\\
2.2	3.2	0.345\\
2.2	3.4	0.335\\
2.2	3.6	0.33\\
2.2	3.8	0.325\\
2.2	4	0.32\\
2.2	4.2	0.31\\
2.2	4.4	0.305\\
2.2	4.6	0.3\\
2.2	4.8	0.295\\
2.2	5	0.29\\
2.2	5.2	0.285\\
2.2	5.4	0.28\\
2.2	5.6	0.275\\
2.2	5.8	0.27\\
2.2	6	0.265\\
2.2	6.2	0.265\\
2.2	6.4	0.26\\
2.2	6.6	0.255\\
2.2	6.8	0.25\\
2.2	7	0.25\\
2.2	7.2	0.245\\
2.4	0.4	0.485\\
2.4	0.6	0.475\\
2.4	0.8	0.465\\
2.4	1	0.45\\
2.4	1.2	0.44\\
2.4	1.4	0.43\\
2.4	1.6	0.42\\
2.4	1.8	0.41\\
2.4	2	0.4\\
2.4	2.2	0.39\\
2.4	2.4	0.38\\
2.4	2.6	0.375\\
2.4	2.8	0.365\\
2.4	3	0.36\\
2.4	3.2	0.35\\
2.4	3.4	0.345\\
2.4	3.6	0.335\\
2.4	3.8	0.33\\
2.4	4	0.325\\
2.4	4.2	0.32\\
2.4	4.4	0.315\\
2.4	4.6	0.31\\
2.4	4.8	0.3\\
2.4	5	0.295\\
2.4	5.2	0.29\\
2.4	5.4	0.29\\
2.4	5.6	0.285\\
2.4	5.8	0.28\\
2.4	6	0.275\\
2.4	6.2	0.27\\
2.4	6.4	0.265\\
2.4	6.6	0.26\\
2.4	6.8	0.26\\
2.4	7	0.255\\
2.4	7.2	0.25\\
2.6	0.4	0.49\\
2.6	0.6	0.475\\
2.6	0.8	0.465\\
2.6	1	0.455\\
2.6	1.2	0.445\\
2.6	1.4	0.435\\
2.6	1.6	0.425\\
2.6	1.8	0.415\\
2.6	2	0.405\\
2.6	2.2	0.395\\
2.6	2.4	0.385\\
2.6	2.6	0.38\\
2.6	2.8	0.37\\
2.6	3	0.365\\
2.6	3.2	0.355\\
2.6	3.4	0.35\\
2.6	3.6	0.345\\
2.6	3.8	0.335\\
2.6	4	0.33\\
2.6	4.2	0.325\\
2.6	4.4	0.32\\
2.6	4.6	0.315\\
2.6	4.8	0.31\\
2.6	5	0.305\\
2.6	5.2	0.3\\
2.6	5.4	0.295\\
2.6	5.6	0.29\\
2.6	5.8	0.285\\
2.6	6	0.28\\
2.6	6.2	0.275\\
2.6	6.4	0.275\\
2.6	6.6	0.27\\
2.6	6.8	0.265\\
2.6	7	0.26\\
2.6	7.2	0.26\\
2.8	0.4	0.49\\
2.8	0.6	0.48\\
2.8	0.8	0.47\\
2.8	1	0.455\\
2.8	1.2	0.445\\
2.8	1.4	0.435\\
2.8	1.6	0.425\\
2.8	1.8	0.42\\
2.8	2	0.41\\
2.8	2.2	0.4\\
2.8	2.4	0.39\\
2.8	2.6	0.385\\
2.8	2.8	0.375\\
2.8	3	0.37\\
2.8	3.2	0.365\\
2.8	3.4	0.355\\
2.8	3.6	0.35\\
2.8	3.8	0.345\\
2.8	4	0.335\\
2.8	4.2	0.33\\
2.8	4.4	0.325\\
2.8	4.6	0.32\\
2.8	4.8	0.315\\
2.8	5	0.31\\
2.8	5.2	0.305\\
2.8	5.4	0.3\\
2.8	5.6	0.295\\
2.8	5.8	0.29\\
2.8	6	0.29\\
2.8	6.2	0.285\\
2.8	6.4	0.28\\
2.8	6.6	0.275\\
2.8	6.8	0.27\\
2.8	7	0.27\\
2.8	7.2	0.265\\
3	0.4	0.49\\
3	0.6	0.48\\
3	0.8	0.47\\
3	1	0.46\\
3	1.2	0.45\\
3	1.4	0.44\\
3	1.6	0.43\\
3	1.8	0.42\\
3	2	0.415\\
3	2.2	0.405\\
3	2.4	0.395\\
3	2.6	0.39\\
3	2.8	0.38\\
3	3	0.375\\
3	3.2	0.37\\
3	3.4	0.36\\
3	3.6	0.355\\
3	3.8	0.35\\
3	4	0.345\\
3	4.2	0.335\\
3	4.4	0.33\\
3	4.6	0.325\\
3	4.8	0.32\\
3	5	0.315\\
3	5.2	0.31\\
3	5.4	0.305\\
3	5.6	0.3\\
3	5.8	0.3\\
3	6	0.295\\
3	6.2	0.29\\
3	6.4	0.285\\
3	6.6	0.28\\
3	6.8	0.28\\
3	7	0.275\\
3	7.2	0.27\\
3.2	0.4	0.49\\
3.2	0.6	0.48\\
3.2	0.8	0.47\\
3.2	1	0.46\\
3.2	1.2	0.45\\
3.2	1.4	0.445\\
3.2	1.6	0.435\\
3.2	1.8	0.425\\
3.2	2	0.415\\
3.2	2.2	0.41\\
3.2	2.4	0.4\\
3.2	2.6	0.395\\
3.2	2.8	0.385\\
3.2	3	0.38\\
3.2	3.2	0.375\\
3.2	3.4	0.365\\
3.2	3.6	0.36\\
3.2	3.8	0.355\\
3.2	4	0.35\\
3.2	4.2	0.345\\
3.2	4.4	0.335\\
3.2	4.6	0.33\\
3.2	4.8	0.325\\
3.2	5	0.32\\
3.2	5.2	0.315\\
3.2	5.4	0.315\\
3.2	5.6	0.31\\
3.2	5.8	0.305\\
3.2	6	0.3\\
3.2	6.2	0.295\\
3.2	6.4	0.29\\
3.2	6.6	0.29\\
3.2	6.8	0.285\\
3.2	7	0.28\\
3.2	7.2	0.275\\
3.4	0.4	0.49\\
3.4	0.6	0.48\\
3.4	0.8	0.475\\
3.4	1	0.465\\
3.4	1.2	0.455\\
3.4	1.4	0.445\\
3.4	1.6	0.435\\
3.4	1.8	0.43\\
3.4	2	0.42\\
3.4	2.2	0.41\\
3.4	2.4	0.405\\
3.4	2.6	0.4\\
3.4	2.8	0.39\\
3.4	3	0.385\\
3.4	3.2	0.375\\
3.4	3.4	0.37\\
3.4	3.6	0.365\\
3.4	3.8	0.36\\
3.4	4	0.355\\
3.4	4.2	0.35\\
3.4	4.4	0.345\\
3.4	4.6	0.335\\
3.4	4.8	0.33\\
3.4	5	0.325\\
3.4	5.2	0.325\\
3.4	5.4	0.32\\
3.4	5.6	0.315\\
3.4	5.8	0.31\\
3.4	6	0.305\\
3.4	6.2	0.3\\
3.4	6.4	0.295\\
3.4	6.6	0.295\\
3.4	6.8	0.29\\
3.4	7	0.285\\
3.4	7.2	0.28\\
3.6	0.4	0.49\\
3.6	0.6	0.485\\
3.6	0.8	0.475\\
3.6	1	0.465\\
3.6	1.2	0.455\\
3.6	1.4	0.45\\
3.6	1.6	0.44\\
3.6	1.8	0.43\\
3.6	2	0.425\\
3.6	2.2	0.415\\
3.6	2.4	0.41\\
3.6	2.6	0.4\\
3.6	2.8	0.395\\
3.6	3	0.39\\
3.6	3.2	0.38\\
3.6	3.4	0.375\\
3.6	3.6	0.37\\
3.6	3.8	0.365\\
3.6	4	0.36\\
3.6	4.2	0.355\\
3.6	4.4	0.345\\
3.6	4.6	0.34\\
3.6	4.8	0.335\\
3.6	5	0.335\\
3.6	5.2	0.33\\
3.6	5.4	0.325\\
3.6	5.6	0.32\\
3.6	5.8	0.315\\
3.6	6	0.31\\
3.6	6.2	0.305\\
3.6	6.4	0.3\\
3.6	6.6	0.3\\
3.6	6.8	0.295\\
3.6	7	0.29\\
3.6	7.2	0.29\\
3.8	0.4	0.49\\
3.8	0.6	0.485\\
3.8	0.8	0.475\\
3.8	1	0.465\\
3.8	1.2	0.46\\
3.8	1.4	0.45\\
3.8	1.6	0.44\\
3.8	1.8	0.435\\
3.8	2	0.425\\
3.8	2.2	0.42\\
3.8	2.4	0.41\\
3.8	2.6	0.405\\
3.8	2.8	0.4\\
3.8	3	0.39\\
3.8	3.2	0.385\\
3.8	3.4	0.38\\
3.8	3.6	0.375\\
3.8	3.8	0.37\\
3.8	4	0.365\\
3.8	4.2	0.355\\
3.8	4.4	0.35\\
3.8	4.6	0.345\\
3.8	4.8	0.34\\
3.8	5	0.335\\
3.8	5.2	0.335\\
3.8	5.4	0.33\\
3.8	5.6	0.325\\
3.8	5.8	0.32\\
3.8	6	0.315\\
3.8	6.2	0.31\\
3.8	6.4	0.31\\
3.8	6.6	0.305\\
3.8	6.8	0.3\\
3.8	7	0.295\\
3.8	7.2	0.295\\
4	0.4	0.49\\
4	0.6	0.485\\
4	0.8	0.475\\
4	1	0.47\\
4	1.2	0.46\\
4	1.4	0.45\\
4	1.6	0.445\\
4	1.8	0.435\\
4	2	0.43\\
4	2.2	0.42\\
4	2.4	0.415\\
4	2.6	0.41\\
4	2.8	0.4\\
4	3	0.395\\
4	3.2	0.39\\
4	3.4	0.385\\
4	3.6	0.38\\
4	3.8	0.37\\
4	4	0.365\\
4	4.2	0.36\\
4	4.4	0.355\\
4	4.6	0.35\\
4	4.8	0.345\\
4	5	0.34\\
4	5.2	0.335\\
4	5.4	0.335\\
4	5.6	0.33\\
4	5.8	0.325\\
4	6	0.32\\
4	6.2	0.315\\
4	6.4	0.31\\
4	6.6	0.31\\
4	6.8	0.305\\
4	7	0.3\\
4	7.2	0.3\\
4.2	0.4	0.495\\
4.2	0.6	0.485\\
4.2	0.8	0.48\\
4.2	1	0.47\\
4.2	1.2	0.46\\
4.2	1.4	0.455\\
4.2	1.6	0.445\\
4.2	1.8	0.44\\
4.2	2	0.43\\
4.2	2.2	0.425\\
4.2	2.4	0.42\\
4.2	2.6	0.41\\
4.2	2.8	0.405\\
4.2	3	0.4\\
4.2	3.2	0.395\\
4.2	3.4	0.385\\
4.2	3.6	0.38\\
4.2	3.8	0.375\\
4.2	4	0.37\\
4.2	4.2	0.365\\
4.2	4.4	0.36\\
4.2	4.6	0.355\\
4.2	4.8	0.35\\
4.2	5	0.345\\
4.2	5.2	0.34\\
4.2	5.4	0.335\\
4.2	5.6	0.335\\
4.2	5.8	0.33\\
4.2	6	0.325\\
4.2	6.2	0.32\\
4.2	6.4	0.315\\
4.2	6.6	0.315\\
4.2	6.8	0.31\\
4.2	7	0.305\\
4.2	7.2	0.3\\
4.4	0.4	0.495\\
4.4	0.6	0.485\\
4.4	0.8	0.48\\
4.4	1	0.47\\
4.4	1.2	0.465\\
4.4	1.4	0.455\\
4.4	1.6	0.45\\
4.4	1.8	0.44\\
4.4	2	0.435\\
4.4	2.2	0.425\\
4.4	2.4	0.42\\
4.4	2.6	0.415\\
4.4	2.8	0.41\\
4.4	3	0.4\\
4.4	3.2	0.395\\
4.4	3.4	0.39\\
4.4	3.6	0.385\\
4.4	3.8	0.38\\
4.4	4	0.375\\
4.4	4.2	0.37\\
4.4	4.4	0.365\\
4.4	4.6	0.36\\
4.4	4.8	0.355\\
4.4	5	0.35\\
4.4	5.2	0.345\\
4.4	5.4	0.34\\
4.4	5.6	0.335\\
4.4	5.8	0.335\\
4.4	6	0.33\\
4.4	6.2	0.325\\
4.4	6.4	0.32\\
4.4	6.6	0.32\\
4.4	6.8	0.315\\
4.4	7	0.31\\
4.4	7.2	0.305\\
4.6	0.4	0.495\\
4.6	0.6	0.485\\
4.6	0.8	0.48\\
4.6	1	0.47\\
4.6	1.2	0.465\\
4.6	1.4	0.455\\
4.6	1.6	0.45\\
4.6	1.8	0.445\\
4.6	2	0.435\\
4.6	2.2	0.43\\
4.6	2.4	0.425\\
4.6	2.6	0.415\\
4.6	2.8	0.41\\
4.6	3	0.405\\
4.6	3.2	0.4\\
4.6	3.4	0.395\\
4.6	3.6	0.39\\
4.6	3.8	0.385\\
4.6	4	0.38\\
4.6	4.2	0.375\\
4.6	4.4	0.37\\
4.6	4.6	0.365\\
4.6	4.8	0.36\\
4.6	5	0.355\\
4.6	5.2	0.35\\
4.6	5.4	0.345\\
4.6	5.6	0.34\\
4.6	5.8	0.335\\
4.6	6	0.335\\
4.6	6.2	0.33\\
4.6	6.4	0.325\\
4.6	6.6	0.32\\
4.6	6.8	0.32\\
4.6	7	0.315\\
4.6	7.2	0.31\\
4.8	0.4	0.495\\
4.8	0.6	0.485\\
4.8	0.8	0.48\\
4.8	1	0.475\\
4.8	1.2	0.465\\
4.8	1.4	0.46\\
4.8	1.6	0.45\\
4.8	1.8	0.445\\
4.8	2	0.44\\
4.8	2.2	0.43\\
4.8	2.4	0.425\\
4.8	2.6	0.42\\
4.8	2.8	0.415\\
4.8	3	0.41\\
4.8	3.2	0.4\\
4.8	3.4	0.395\\
4.8	3.6	0.39\\
4.8	3.8	0.385\\
4.8	4	0.38\\
4.8	4.2	0.375\\
4.8	4.4	0.37\\
4.8	4.6	0.365\\
4.8	4.8	0.36\\
4.8	5	0.36\\
4.8	5.2	0.355\\
4.8	5.4	0.35\\
4.8	5.6	0.345\\
4.8	5.8	0.34\\
4.8	6	0.335\\
4.8	6.2	0.335\\
4.8	6.4	0.33\\
4.8	6.6	0.325\\
4.8	6.8	0.32\\
4.8	7	0.32\\
4.8	7.2	0.315\\
5	0.4	0.495\\
5	0.6	0.49\\
5	0.8	0.48\\
5	1	0.475\\
5	1.2	0.465\\
5	1.4	0.46\\
5	1.6	0.455\\
5	1.8	0.445\\
5	2	0.44\\
5	2.2	0.435\\
5	2.4	0.43\\
5	2.6	0.42\\
5	2.8	0.415\\
5	3	0.41\\
5	3.2	0.405\\
5	3.4	0.4\\
5	3.6	0.395\\
5	3.8	0.39\\
5	4	0.385\\
5	4.2	0.38\\
5	4.4	0.375\\
5	4.6	0.37\\
5	4.8	0.365\\
5	5	0.36\\
5	5.2	0.355\\
5	5.4	0.355\\
5	5.6	0.35\\
5	5.8	0.345\\
5	6	0.34\\
5	6.2	0.335\\
5	6.4	0.335\\
5	6.6	0.33\\
5	6.8	0.325\\
5	7	0.325\\
5	7.2	0.32\\
5.2	0.4	0.495\\
5.2	0.6	0.49\\
5.2	0.8	0.48\\
5.2	1	0.475\\
5.2	1.2	0.47\\
5.2	1.4	0.46\\
5.2	1.6	0.455\\
5.2	1.8	0.45\\
5.2	2	0.44\\
5.2	2.2	0.435\\
5.2	2.4	0.43\\
5.2	2.6	0.425\\
5.2	2.8	0.42\\
5.2	3	0.415\\
5.2	3.2	0.41\\
5.2	3.4	0.4\\
5.2	3.6	0.395\\
5.2	3.8	0.39\\
5.2	4	0.385\\
5.2	4.2	0.385\\
5.2	4.4	0.38\\
5.2	4.6	0.375\\
5.2	4.8	0.37\\
5.2	5	0.365\\
5.2	5.2	0.36\\
5.2	5.4	0.355\\
5.2	5.6	0.35\\
5.2	5.8	0.35\\
5.2	6	0.345\\
5.2	6.2	0.34\\
5.2	6.4	0.335\\
5.2	6.6	0.335\\
5.2	6.8	0.33\\
5.2	7	0.325\\
5.2	7.2	0.325\\
5.4	0.4	0.495\\
5.4	0.6	0.49\\
5.4	0.8	0.48\\
5.4	1	0.475\\
5.4	1.2	0.47\\
5.4	1.4	0.465\\
5.4	1.6	0.455\\
5.4	1.8	0.45\\
5.4	2	0.445\\
5.4	2.2	0.44\\
5.4	2.4	0.43\\
5.4	2.6	0.425\\
5.4	2.8	0.42\\
5.4	3	0.415\\
5.4	3.2	0.41\\
5.4	3.4	0.405\\
5.4	3.6	0.4\\
5.4	3.8	0.395\\
5.4	4	0.39\\
5.4	4.2	0.385\\
5.4	4.4	0.38\\
5.4	4.6	0.375\\
5.4	4.8	0.37\\
5.4	5	0.37\\
5.4	5.2	0.365\\
5.4	5.4	0.36\\
5.4	5.6	0.355\\
5.4	5.8	0.35\\
5.4	6	0.35\\
5.4	6.2	0.345\\
5.4	6.4	0.34\\
5.4	6.6	0.335\\
5.4	6.8	0.335\\
5.4	7	0.33\\
5.4	7.2	0.325\\
5.6	0.4	0.495\\
5.6	0.6	0.49\\
5.6	0.8	0.485\\
5.6	1	0.475\\
5.6	1.2	0.47\\
5.6	1.4	0.465\\
5.6	1.6	0.46\\
5.6	1.8	0.45\\
5.6	2	0.445\\
5.6	2.2	0.44\\
5.6	2.4	0.435\\
5.6	2.6	0.43\\
5.6	2.8	0.425\\
5.6	3	0.42\\
5.6	3.2	0.415\\
5.6	3.4	0.41\\
5.6	3.6	0.405\\
5.6	3.8	0.4\\
5.6	4	0.395\\
5.6	4.2	0.39\\
5.6	4.4	0.385\\
5.6	4.6	0.38\\
5.6	4.8	0.375\\
5.6	5	0.37\\
5.6	5.2	0.365\\
5.6	5.4	0.365\\
5.6	5.6	0.36\\
5.6	5.8	0.355\\
5.6	6	0.35\\
5.6	6.2	0.35\\
5.6	6.4	0.345\\
5.6	6.6	0.34\\
5.6	6.8	0.335\\
5.6	7	0.335\\
5.6	7.2	0.33\\
5.8	0.4	0.495\\
5.8	0.6	0.49\\
5.8	0.8	0.485\\
5.8	1	0.475\\
5.8	1.2	0.47\\
5.8	1.4	0.465\\
5.8	1.6	0.46\\
5.8	1.8	0.455\\
5.8	2	0.445\\
5.8	2.2	0.44\\
5.8	2.4	0.435\\
5.8	2.6	0.43\\
5.8	2.8	0.425\\
5.8	3	0.42\\
5.8	3.2	0.415\\
5.8	3.4	0.41\\
5.8	3.6	0.405\\
5.8	3.8	0.4\\
5.8	4	0.395\\
5.8	4.2	0.39\\
5.8	4.4	0.385\\
5.8	4.6	0.38\\
5.8	4.8	0.38\\
5.8	5	0.375\\
5.8	5.2	0.37\\
5.8	5.4	0.365\\
5.8	5.6	0.36\\
5.8	5.8	0.36\\
5.8	6	0.355\\
5.8	6.2	0.35\\
5.8	6.4	0.345\\
5.8	6.6	0.345\\
5.8	6.8	0.34\\
5.8	7	0.335\\
5.8	7.2	0.335\\
6	0.4	0.495\\
6	0.6	0.49\\
6	0.8	0.485\\
6	1	0.48\\
6	1.2	0.47\\
6	1.4	0.465\\
6	1.6	0.46\\
6	1.8	0.455\\
6	2	0.45\\
6	2.2	0.445\\
6	2.4	0.44\\
6	2.6	0.43\\
6	2.8	0.425\\
6	3	0.42\\
6	3.2	0.415\\
6	3.4	0.41\\
6	3.6	0.405\\
6	3.8	0.405\\
6	4	0.4\\
6	4.2	0.395\\
6	4.4	0.39\\
6	4.6	0.385\\
6	4.8	0.38\\
6	5	0.375\\
6	5.2	0.375\\
6	5.4	0.37\\
6	5.6	0.365\\
6	5.8	0.36\\
6	6	0.36\\
6	6.2	0.355\\
6	6.4	0.35\\
6	6.6	0.345\\
6	6.8	0.345\\
6	7	0.34\\
6	7.2	0.335\\
6.2	0.4	0.495\\
6.2	0.6	0.49\\
6.2	0.8	0.485\\
6.2	1	0.48\\
6.2	1.2	0.475\\
6.2	1.4	0.465\\
6.2	1.6	0.46\\
6.2	1.8	0.455\\
6.2	2	0.45\\
6.2	2.2	0.445\\
6.2	2.4	0.44\\
6.2	2.6	0.435\\
6.2	2.8	0.43\\
6.2	3	0.425\\
6.2	3.2	0.42\\
6.2	3.4	0.415\\
6.2	3.6	0.41\\
6.2	3.8	0.405\\
6.2	4	0.4\\
6.2	4.2	0.395\\
6.2	4.4	0.39\\
6.2	4.6	0.39\\
6.2	4.8	0.385\\
6.2	5	0.38\\
6.2	5.2	0.375\\
6.2	5.4	0.37\\
6.2	5.6	0.37\\
6.2	5.8	0.365\\
6.2	6	0.36\\
6.2	6.2	0.355\\
6.2	6.4	0.355\\
6.2	6.6	0.35\\
6.2	6.8	0.345\\
6.2	7	0.345\\
6.2	7.2	0.34\\
6.4	0.4	0.495\\
6.4	0.6	0.49\\
6.4	0.8	0.485\\
6.4	1	0.48\\
6.4	1.2	0.475\\
6.4	1.4	0.47\\
6.4	1.6	0.46\\
6.4	1.8	0.455\\
6.4	2	0.45\\
6.4	2.2	0.445\\
6.4	2.4	0.44\\
6.4	2.6	0.435\\
6.4	2.8	0.43\\
6.4	3	0.425\\
6.4	3.2	0.42\\
6.4	3.4	0.415\\
6.4	3.6	0.41\\
6.4	3.8	0.405\\
6.4	4	0.405\\
6.4	4.2	0.4\\
6.4	4.4	0.395\\
6.4	4.6	0.39\\
6.4	4.8	0.385\\
6.4	5	0.38\\
6.4	5.2	0.38\\
6.4	5.4	0.375\\
6.4	5.6	0.37\\
6.4	5.8	0.365\\
6.4	6	0.365\\
6.4	6.2	0.36\\
6.4	6.4	0.355\\
6.4	6.6	0.355\\
6.4	6.8	0.35\\
6.4	7	0.345\\
6.4	7.2	0.345\\
6.6	0.4	0.495\\
6.6	0.6	0.49\\
6.6	0.8	0.485\\
6.6	1	0.48\\
6.6	1.2	0.475\\
6.6	1.4	0.47\\
6.6	1.6	0.465\\
6.6	1.8	0.46\\
6.6	2	0.455\\
6.6	2.2	0.445\\
6.6	2.4	0.44\\
6.6	2.6	0.435\\
6.6	2.8	0.43\\
6.6	3	0.43\\
6.6	3.2	0.425\\
6.6	3.4	0.42\\
6.6	3.6	0.415\\
6.6	3.8	0.41\\
6.6	4	0.405\\
6.6	4.2	0.4\\
6.6	4.4	0.395\\
6.6	4.6	0.39\\
6.6	4.8	0.39\\
6.6	5	0.385\\
6.6	5.2	0.38\\
6.6	5.4	0.375\\
6.6	5.6	0.375\\
6.6	5.8	0.37\\
6.6	6	0.365\\
6.6	6.2	0.365\\
6.6	6.4	0.36\\
6.6	6.6	0.355\\
6.6	6.8	0.355\\
6.6	7	0.35\\
6.6	7.2	0.345\\
6.8	0.4	0.495\\
6.8	0.6	0.49\\
6.8	0.8	0.485\\
6.8	1	0.48\\
6.8	1.2	0.475\\
6.8	1.4	0.47\\
6.8	1.6	0.465\\
6.8	1.8	0.46\\
6.8	2	0.455\\
6.8	2.2	0.45\\
6.8	2.4	0.445\\
6.8	2.6	0.44\\
6.8	2.8	0.435\\
6.8	3	0.43\\
6.8	3.2	0.425\\
6.8	3.4	0.42\\
6.8	3.6	0.415\\
6.8	3.8	0.41\\
6.8	4	0.405\\
6.8	4.2	0.405\\
6.8	4.4	0.4\\
6.8	4.6	0.395\\
6.8	4.8	0.39\\
6.8	5	0.385\\
6.8	5.2	0.385\\
6.8	5.4	0.38\\
6.8	5.6	0.375\\
6.8	5.8	0.37\\
6.8	6	0.37\\
6.8	6.2	0.365\\
6.8	6.4	0.36\\
6.8	6.6	0.36\\
6.8	6.8	0.355\\
6.8	7	0.35\\
6.8	7.2	0.35\\
7	0.4	0.495\\
7	0.6	0.49\\
7	0.8	0.485\\
7	1	0.48\\
7	1.2	0.475\\
7	1.4	0.47\\
7	1.6	0.465\\
7	1.8	0.46\\
7	2	0.455\\
7	2.2	0.45\\
7	2.4	0.445\\
7	2.6	0.44\\
7	2.8	0.435\\
7	3	0.43\\
7	3.2	0.425\\
7	3.4	0.42\\
7	3.6	0.415\\
7	3.8	0.415\\
7	4	0.41\\
7	4.2	0.405\\
7	4.4	0.4\\
7	4.6	0.395\\
7	4.8	0.395\\
7	5	0.39\\
7	5.2	0.385\\
7	5.4	0.38\\
7	5.6	0.38\\
7	5.8	0.375\\
7	6	0.37\\
7	6.2	0.37\\
7	6.4	0.365\\
7	6.6	0.36\\
7	6.8	0.36\\
7	7	0.355\\
7	7.2	0.35\\
};

\end{axis}

\end{tikzpicture}%